\documentclass[twocolumn]{aastex631}

\usepackage{soul}

\usepackage{CJK}
\usepackage{amsmath}

\begin{document}
\title{The Relation between AGN and Host Galaxy Properties in the JWST Era: \\
I. Seyferts at Cosmic Noon are Obscured and Disturbed}

\correspondingauthor{Nina Bonaventura}
\email{nbonaventura@arizona.edu}

\author[0000-0001-8470-7094]{Nina Bonaventura}
\affiliation{Steward Observatory, University of Arizona, 933 North Cherry Avenue, Tucson, AZ 85721, USA}

\author[0000-0002-6221-1829]{Jianwei Lyu (\begin{CJK}{UTF8}{gbsn}吕建伟\end{CJK})}
\affiliation{Steward Observatory, University of Arizona,
933 North Cherry Avenue, Tucson, AZ 85721, USA}

\author[0000-0003-2303-6519]{George H. Rieke}
\affiliation{Steward Observatory, University of Arizona,
933 North Cherry Avenue, Tucson, AZ 85721, USA}

\author[0000-0002-8909-8782]{Stacey Alberts}
\affiliation{Steward Observatory, University of Arizona,
933 North Cherry Avenue, Tucson, AZ 85721, USA}

\author[0000-0001-9262-9997]{Christopher N. A. Willmer}
\affiliation{Steward Observatory, University of Arizona, 933 North Cherry Avenue, Tucson, AZ 85721, USA}

\author[0000-0003-4528-5639]{Pablo G. P\'erez-Gonz\'alez}
\affiliation{Centro de Astrobiolog\'{\i}a (CAB), CSIC-INTA, Ctra. de Ajalvir km 4, Torrej\'on de Ardoz, E-28850, Madrid, Spain}

\author[0000-0002-8651-9879]{Andrew J. Bunker} 
\affiliation{Department of Physics, University of Oxford, Denys Wilkinson Building, Keble Road, Oxford OX1 3RH, UK}

\author[0000-0002-3402-6917]{Meredith Stone}
\affiliation{Steward Observatory, University of Arizona, 933 North Cherry Avenue, Tucson, AZ 85721, USA}

\author[0000-0003-2388-8172]{Francesco D'Eugenio}
\affiliation{Kavli Institute for Cosmology, University of Cambridge, Madingley Road, Cambridge, CB3 0HA, UK}
\affiliation{Cavendish Laboratory, University of Cambridge, 19 JJ Thomson Avenue, Cambridge, CB3 0HE, UK}

\author[0000-0003-2919-7495]{Christina C.\ Williams}
\affiliation{NSF's National Optical-Infrared Astronomy Research Laboratory, 950 North Cherry Avenue, Tucson, AZ 85719, USA}
\affiliation{Steward Observatory, University of Arizona, 933 North Cherry Avenue, Tucson, AZ 85721, USA}

\author[0000-0003-0695-4414]{Michael V. Maseda}
\affiliation{Department of Astronomy, University of Wisconsin-Madison, 475 N. Charter St., Madison, WI 53706 USA}

\author[0000-0002-4201-7367]{Chris J. Willott}
\affil{NRC Herzberg, 5071 West Saanich Rd, Victoria, BC V9E 2E7, Canada}

\author[0000-0001-7673-2257]{Zhiyuan Ji}
\affiliation{Steward Observatory, University of Arizona, 933 N. Cherry Ave, Tucson, AZ 85721, USA}

\author[0000-0003-0215-1104]{William M. Baker}
\affiliation{Kavli Institute for Cosmology, University of Cambridge, Madingley Road, Cambridge CB3 0HA, UK} 
\affiliation{Cavendish Laboratory, University of Cambridge, 19 JJ Thomson Avenue, Cambridge CB3 0HE, UK}

\author[0000-0002-6719-380X]{Stefano Carniani}
\affiliation{Scuola Normale Superiore, Piazza dei Cavalieri 7, I-56126 Pisa, Italy}

\author[0000-0003-3458-2275]{Stephane Charlot}
\affiliation{Sorbonne Universit\'e, UPMC-CNRS, UMR7095, Institut d'Astrophysique de Paris, F-75014 Paris, France}

\author[0000-0002-7636-0534]{Jacopo Chevallard}
\affiliation{Department of Physics, University of Oxford, Denys Wilkinson Building, Keble Road, Oxford OX1 3RH, UK}

\author[0000-0002-9551-0534]{Emma Curtis-Lake}
\affiliation{Centre for Astrophysics Research, Department of Physics, Astronomy and Mathematics, University of Hertfordshire, Hatfield AL10 9AB, UK}

\author[0000-0002-2929-3121]{Daniel J. Eisenstein}
\affiliation{Center for Astrophysics | Harvard and Smithsonian, 60 Garden Street, Cambridge, MA 02138, USA}

\author[0000-0003-4565-8239]{Kevin Hainline}
\affiliation{Steward Observatory, University of Arizona, 933 N. Cherry Ave, Tucson, AZ 85721, USA}

\author[0000-0002-8543-761X]{Ryan Hausen}
\affiliation{Department of Physics and Astronomy, The Johns Hopkins University, 3400 N. Charles St., Baltimore, MD 21218, USA}

\author[0000-0002-7524-374X]{Erica J.\ Nelson}
\affiliation{Department for Astrophysical and Planetary Science, University of Colorado, Boulder, CO 80309, USA}

\author[0000-0002-7893-6170]{Marcia J. Rieke}
\affiliation{Steward Observatory, University of Arizona, 933 N. Cherry Ave, Tucson, AZ 85721, USA}

\author[0000-0002-4271-0364]{Brant Robertson}
\affiliation{Department of Astronomy and Astrophysics, University of California, Santa Cruz, 1156 High Street, Santa Cruz, CA 95064, USA}

\author[0000-0003-4702-7561]{Irene Shivaei}
\affiliation{Centro de Astrobiolog\'{\i}a (CAB), CSIC-INTA, Ctra. de Ajalvir km 4, Torrej\'on de Ardoz, E-28850, Madrid, Spain}

\begin{abstract}

The morphology of a galaxy reflects the mix of physical processes occurring within and around it, offering indirect clues to its formation and evolution. We apply both visual classification and computer vision to test the suspected connection between galaxy mergers and AGN activity, as evidenced by a close/merging galaxy pair, or tidal features surrounding an apparently singular system. We use JADES JWST/NIRCam imagery of a complete, mutliwavelength AGN sample recently expanded with JWST/MIRI photometry.  This 0.9-25 $\mu$m dataset enables constraints on the host galaxy morphologies of a broad range of AGN beyond z$\sim$1, including heavily obscured examples missing from previous studies. Our primary AGN sample consists of 243 lightly to highly obscured X-ray-selected AGN and 138 presumed Compton-thick, mid-infrared-bright/X-ray-faint AGN revealed by MIRI. Utilizing the shape asymmetry morphology indicator, $A_S$, as the metric for disturbance, we find that 88\% of the Seyferts sampled are strongly spatially disturbed ($A_S>0.2$). The experimental design we employ reveals a $\gtrsim 3\sigma$ obscuration-merger ($N_H$-$A_S$) correlation at $0.6<z<2.4$, and also recovers a physical distinction between the X-ray- and mid-IR-detected AGN suggestive of their link to a common evolutionary scenario. Placing the observed pattern of disturbances in the context of the other average host galaxy properties, we conclude that mergers are common amongst obscured AGN. This finding presents tension with the leading model on AGN fueling that requires Seyfert AGN with sub-quasar luminosities ($L_{bol} < 10^{45}$ ergs/s) to evolve only through non-merger mechanisms.
\end{abstract}

\section{Introduction}
In the hierarchical framework for large-scale structure formation in the Universe, a `cosmic cycle' for galaxy formation and evolution is expected as a direct result of regularly occurring galaxy mergers \citep{Hopkins2006a}. In this picture, two gas-rich disk galaxies of similar masses merge to violently drive a nuclear inflow of gas towards the center of the system that triggers both starbursts and active galactic nuclei (AGN), and therefore the simultaneous growth of black holes and stellar bulges in the centers of galaxies. There is significant observational support of this modeled co-evolution of galaxies with their central supermassive black holes (SMBHs) in quasars ($L_{bol} \ge 10^{45}$ ergs/s), given their extreme luminosities and relative ease of detection in comparison to Seyfert AGN with lower luminosities and stronger relative contamination from their host galaxies (see, e.g., \citealp{Hopkins2006b} and references therein). As a result, the historical record of quasar observations helped pave the way to the discovery of numerous galaxy scaling relations, most famously the empirical $M-\sigma_{disp}$ relation showing the tight correlation between black hole mass and host galaxy velocity dispersion, clearly demonstrating a link between the origin of galaxies and SMBHs (see, e.g., \citealp{KH2000, DSH2005}). There is also the persistent coincidence of intensive star formation with embedded active nuclei in ultra-luminous infrared galaxies (ULIRGs) commonly associated with mergers \citep{Sanders1996, Jogee2006}. 

It is predicted that the quasar luminosity can only be triggered when the galaxies involved in gas-rich, major mergers have a dark matter halo mass above a critical value, lying between $10^{12}$ and $10^{13}$ $M_{\odot}$ (and corresponding black hole mass, $M_{BH}$ $\gtrsim$ $10^{7}$ $M_{\odot}$, \citealp{Hickox2009}). Their subsequent evolution is then expected to proceed as follows: the active nucleus is initially obscured at optical and soft X-ray wavelengths by a thick cocoon of gas and dust, for the majority of the merger lifetime, where the obscuring material can be confined to the nuclear region of the galaxy, as well as distributed to galaxy-wide scales by the triggering merger event \citep{Hickox2018}. This describes the early obscured, Type II phase of a quasar, that ends when the nuclear radiation grows powerful enough to eject the obscuring material, revealing the relatively brief, luminous, unobscured Type I quasar phase. At this point, when the AGN can no longer be sustained, it is predicted that a merger remnant emerges in the form of a normal, passively evolving galaxy with a central stellar bulge and SMBH, typically in the form of a spheroid. As such, conventional wisdom has held that Seyferts, that do not require the violence of a major gas-rich merger to fuel their lower luminosities, and that typically present with a disk morphology, must be passively and stochastically fueled, explicitly by non-merger mechanisms \citep{KK2004, Hopkins2006a, Hopkins2014}.    

However, the observational evidence to distinguish between these two scenarios for SMBH growth -- either confined to the quasar regime in short and dramatic episodes \citep{Hickox2018, Hopkins2008, Gilli2007}, or secular and stochastic evolution \citep{KK2004, Hopkins2006a, Hopkins2014} - is ambiguous \citep{Treister2012,Ji2022}, and the issue of their relative roles remains unresolved \citep[see, e.g.,][]{Shah2020, Pierce2023, Hernandez2023, Villforth2023}. A further observational complexity is that Type I quasar host galaxies tend to lie on the main sequence \citep[e.g.,][]{Xu2015,Zhang2016} rather than emerging from ULIRGs or lying in post-starburst, passively evolving galaxies, as might be expected from theoretical predictions. Finally, it is not clear if this popular model for quasar evolution and its exclusion of AGN at lower luminosities and black hole masses \citep{Hopkins2014} is born out in the growing body of observational evidence \citep{Kocevski2015,Donley2012,Treister2012}.

Clues to the formation and evolution of AGN should lie in their host galaxy morphologies; there have been many programs to image the host galaxies of nearby and moderate redshift AGN and quasars \citep[e.g.,][]{Kauffmann2003, Cisternas2011,  Kim2017}. Although there are disagreements in the assignment of disturbances in individual cases, in general there is broad consistency among the studies that roughly 20 - 30\% of the hosts of nearby AGN appear disturbed, as illustrated in the comprehensive source-by-source summary by \citet{Kim2017}; see Section~\ref{calibration}. This agreement holds when different techniques are used, providing credibility to the assignments.

The great majority of the available studies have focused largely on Type I unobscured or lightly obscured AGN.  However, the discovery of substantial numbers of Type II AGN in the SDSS \citep{Zakamska2004} has enabled similar observations of their host galaxies. For example, \citet{Urbano2019} report that, in a sample of 41 Type II quasars, $34^{+6}_{-9}$\% of the host galaxies are highly disturbed; they summarize the existing work as showing that $44 -62$\% of Type II hosts are disturbed. \citet{Zakamska2019} present \textit{Hubble Space Telescope} (HST) images of 10 extremely dust-reddened (`red') quasars (ERQs)\footnote{\citet{Zakamska2019} suggest that the relatively low number of disturbed hosts detected amongst the ERQs of their study may be due to the undetectability of faint merger signatures in the HST imagery utilized, e.g. if these systems are late-stage merger remnants and/or lie in very dusty host galaxies.} and 6 Type II AGN, finding, respectively, that $2/10$ and $2/6$ of the hosts are disturbed. \citet{Glikman2024} and references therein report on a spectroscopically confirmed sample of $\sim130$ Type I, moderately obscured red quasars detected between $0.1 < z < 3$ (\citealp{Glikman2004, Glikman2007, Urrutia2009, Glikman2012, Glikman2013}) that largely appear in mergers ($>80$\% ) in HST imagery (\citealp{Urrutia2008,Glikman2015}), and likely represent a transitional phase in the merger-driven process coinciding with the start of the `blow-out' phase. That is, it appears that these obscured AGN also show a significant fraction of disturbed hosts, as classified by the available methods. However, even the Type II samples discussed above do not include the most heavily obscured, Compton-thick ($N_{H} \ge 10^{24}$  $cm^{-2}$) AGN. A comprehensive study of the relation of host morphology to AGN obscuration needs to draw on the full range of AGN identification methods (e.g., \citealp{Lyu2022}), including those centered in the infrared that can detect AGN that are so obscured that their traces are hidden in other wavelength ranges (e.g., \citealp{Lyu2024a}). 

The high sensitivity, high resolution capabilities of NIRCam open multiple possibilities to study AGN host galaxy morphology at Cosmic Noon; this will enable many studies of host galaxy behavior and provides the motivation for this paper. We also build on the thoroughness of the AGN identification in the GOODS-South (GOODS-S) field \citep{Lyu2022}, including the identification of a large sample of previously unknown obscured AGN out to high redshift with the JWST Mid-Infrared Instrument(MIRI) \citep{Lyu2024a}. Together, these advances allow for the study of the connection between large-scale morphological disturbances in an active galaxy and evidence of obscuration by dust and/or gas surrounding its AGN.  Here (Paper I), we present the details of our analysis of the JWST observations. In Paper II, we will supplement Paper I with additional datasets and analysis to support an evolutionary scenario for massive Seyfert galaxies implied by the results of the initial study.

Our discussion will develop as follows. Section 2 provides an overview of the new data we have used, specifically JWST/NIRCam images \citep{RiekeM2023} of the X-ray-selected subset of AGN from the pre-JWST multiwavelength sample contained \citep{Lyu2022}, as well as a new sample selected with JWST/MIRI. We also summarize the AGN identification methods relevant to our study. Section 3 describes the methodology of our analysis. Section 4 describes the design of our study and Section 5 presents the results, which are discussed in Section 6. In Section 7 we summarize and conclude the paper.

The galaxy parameter values quoted in this paper are drawn directly from the SED analyses of \citet{Lyu2022} and \citet{Lyu2024a}, which adopt a slighty different cosmology:  
\citet{Lyu2022} assumes $\Omega _m=0.27$ and $H_0 = 71$ km~s$^{-1}$Mpc$^{-1}$, while \citep{Lyu2024a} assumes $\Omega _m = 0.287$ and $H_0=69.3$ km~s$^{-1}$Mpc$^{-1}$. This minor difference is insignificant with regard to our conclusions.

  \begin{figure}[h!]
     \centering
         \centering
         \includegraphics[width=1\linewidth]{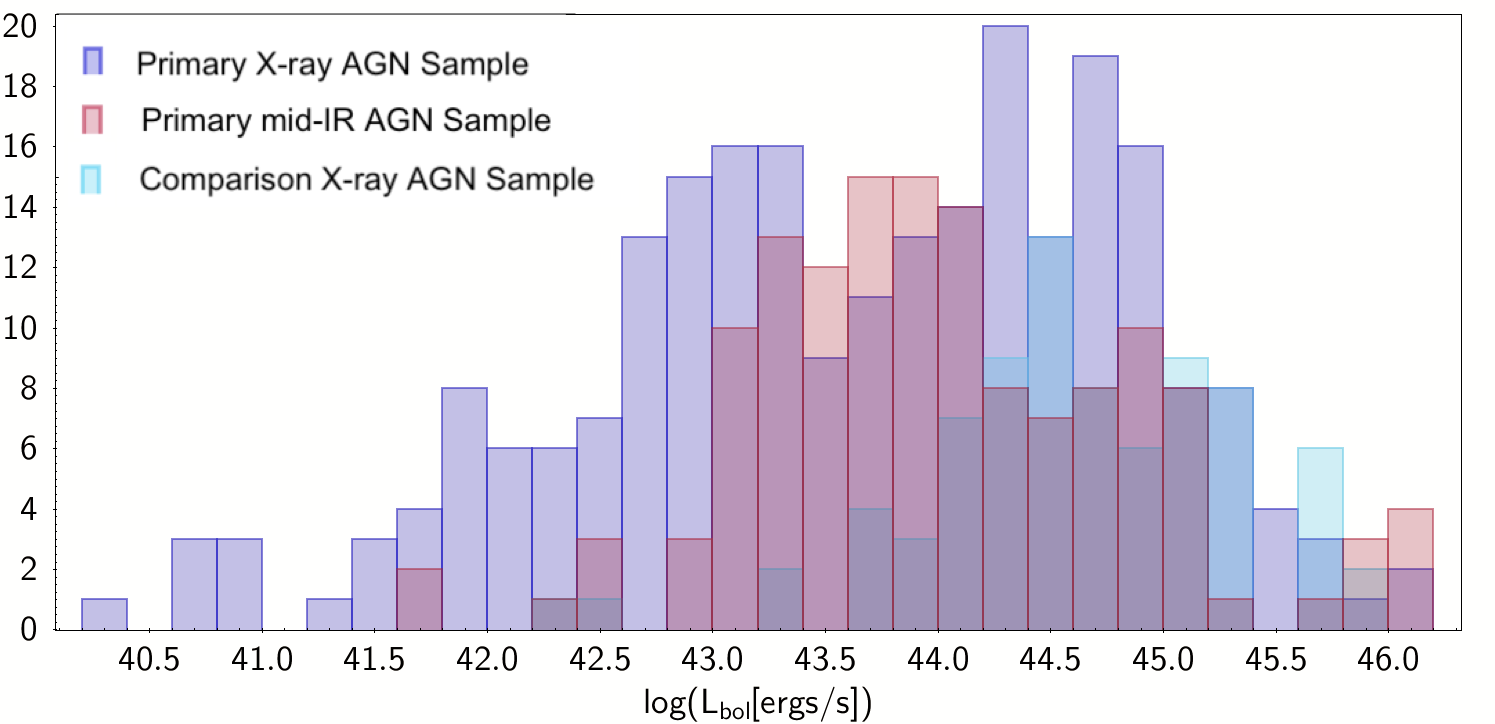}
     \hfill     
         \centering
         \includegraphics[width=1\linewidth]{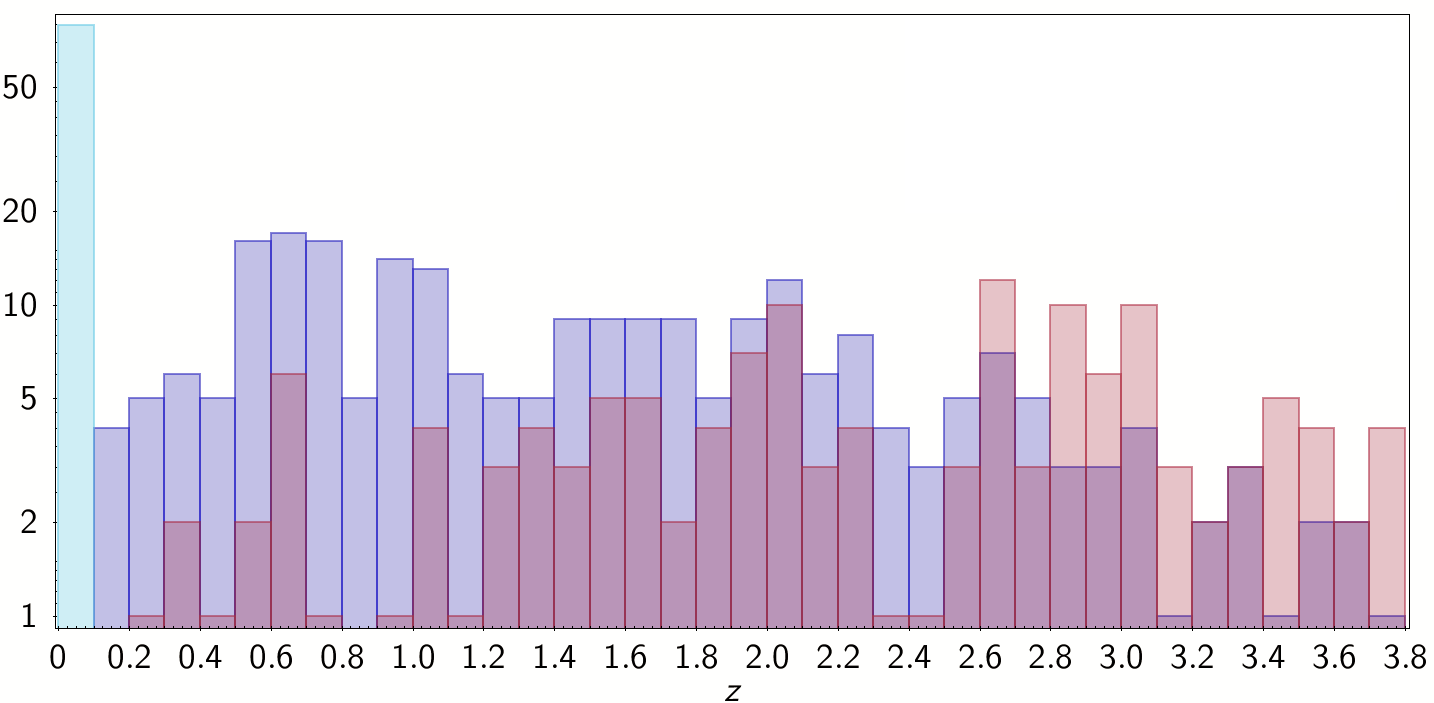}
     \hfill
         \centering
         \includegraphics[width=1\linewidth]{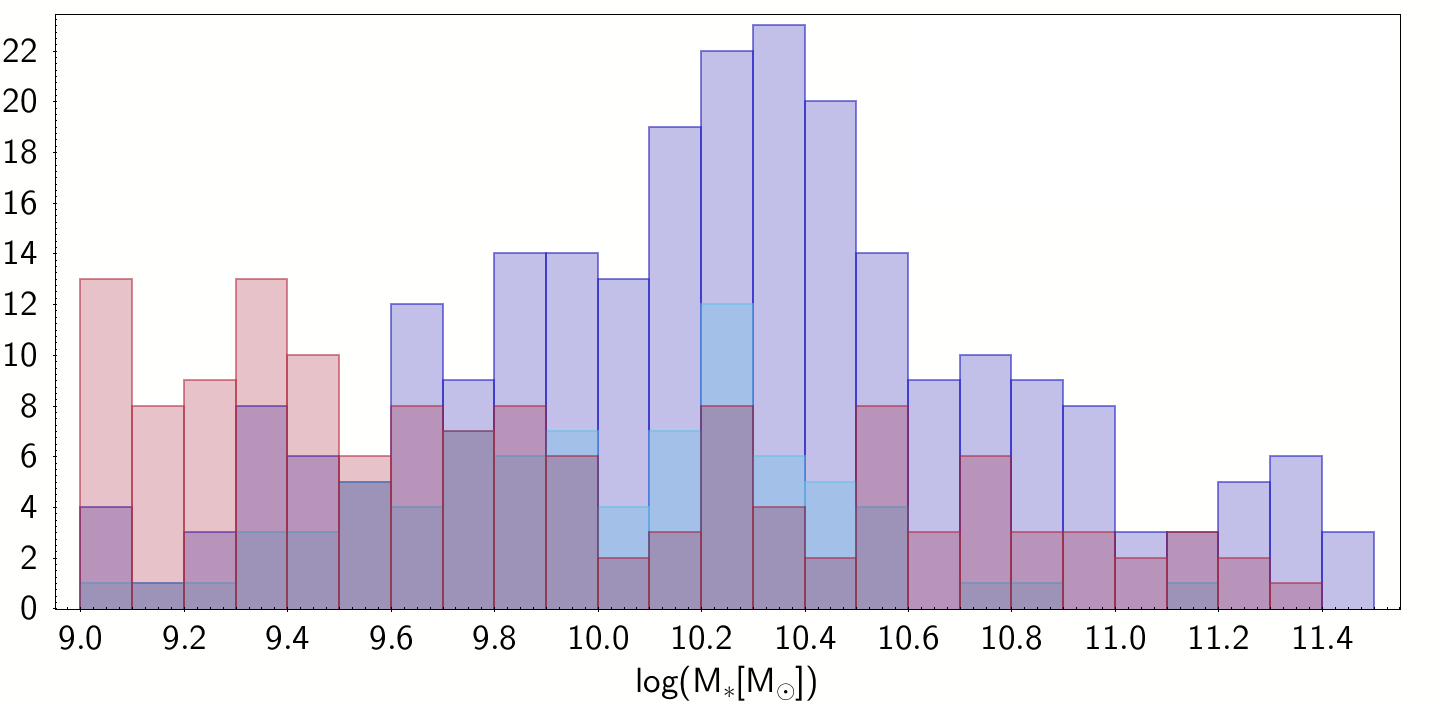}
     \hfill
         \centering
         \includegraphics[width=1\linewidth]{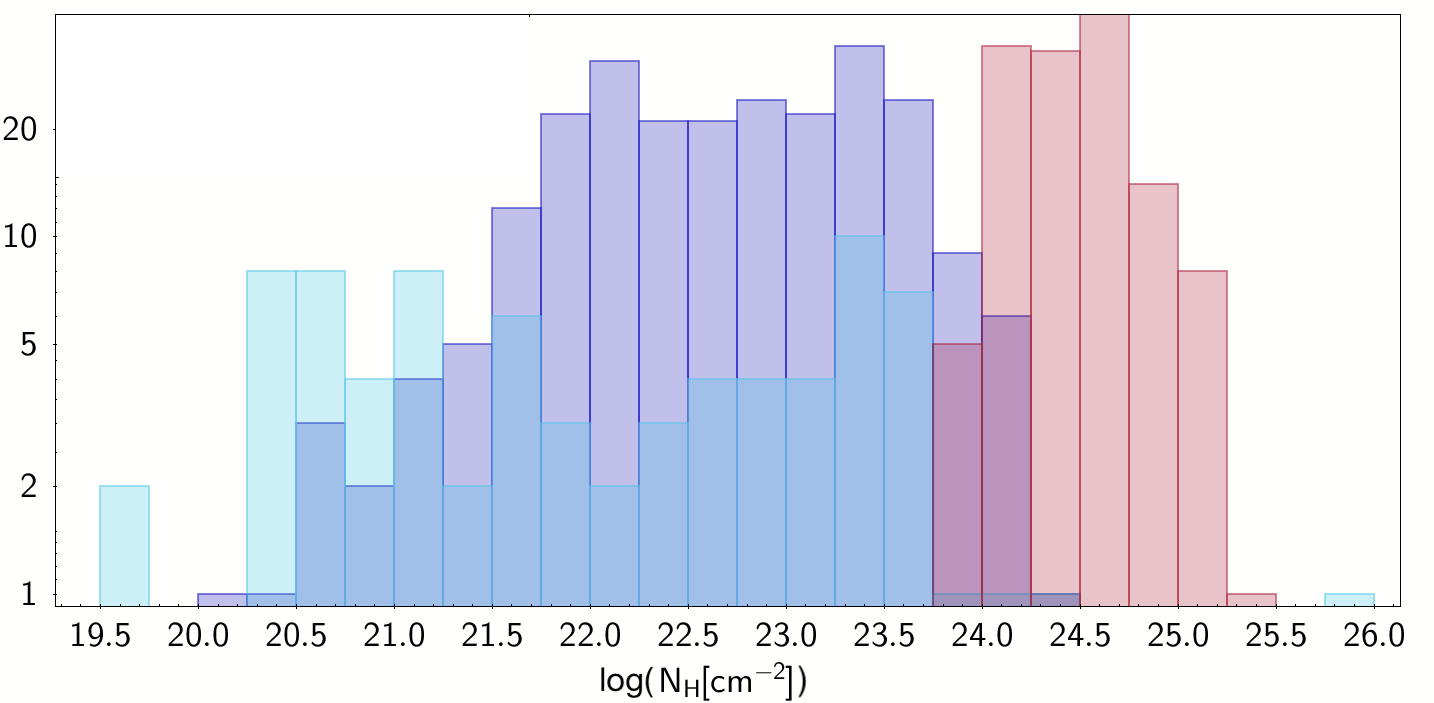}
         
         \caption{Histograms showing the distribution of the bolometric luminosities ($L_{bol}$), redshifts (\textit{z}), stellar masses ($M_{*}$), and the hydrogen absorbing column densities ($N_{H}$) of the entire AGN sample (full primary sample plus local comparison sample) (\citealp{Lyu2022,Lyu2024a,Lyu2024b,Tueller2008}). The $N_H$ values for the mid-IR-bright/X-ray-faint (i.e., Compton-thick) AGN sample were simulated as described in Section ~\ref{sec: sample_data}.}. 
\label{fig: full_sample_properties_hist}
\end{figure}

\section{Samples and Data}

This study is enabled by previous work to identify AGN in the GOODS-S region using ultra-deep \emph{Chandra} and \emph{Spitzer} imaging, now complemented by both the mid-IR identification of obscured AGN in the same field with JWST/MIRI, plus the ability to examine the host galaxies uniformly using JWST/NIRCam. 

\subsection{Primary AGN Sample}

Our primary AGN sample consists of a subset of the X-ray-identified (0.5-7 keV band) sources contained in the comprehensive census of GOODS-S/HUDF AGN analyzed in \citet{Lyu2022}, which also contains AGN detected at optical, infrared, and/or radio wavelengths over an area spanning $\sim$ 170 arcmin$^2$; as well as the the X-ray-identified subset identified in GOODS-N using the identical methodology \citep{Lyu2024b}. We also incorporate a new sample of JWST/MIRI-detected AGN revealed by the Systematic Mid-infrared Instrument Legacy Extragalactic Survey (SMILES, \citealp{Alberts2024, RiekeG2024}), a JWST Cycle 1 GTO program (PI: G. Rieke) that has targeted the central 35 $arcmin^2$ region of GOODS-S with eight MIRI filters from 5.6 to 25.5 $\mu$m \citep{Lyu2024a}. Both Type I and II AGN are included in these samples, detected out to $z \sim 6$ via spectroscopic and/or photometric measurements, and are presented with a full set of SED-derived physical parameters. 

In our analysis, we include the sources lying at $0.1<z<5$, but focus mainly on those restricted to $0.6<z<2.4$ that are matched on numerous galaxy properties and therefore enable valid comparisons.

\subsubsection{\textit{Chandra} (0.5-7 keV) X-ray AGN Subsample}

The X-ray AGN subset of the primary sample was defined as all sources in the catalog of \citet{Lyu2022} and \citet{Lyu2024b} for which the ratio of the 0.5-7 keV X-ray luminosity to the 3 GHz radio luminosity exceeds the physical threshold for stellar processes in a galaxy ($L_{X,int}$ [ergs/s]/$L_{3GHz}$ [W/Hz]$>8x10^{18}$, \citealp{Lyu2022}), and is therefore confidently attributed to AGN emission \citep{Alberts2020}. In fact, \citet{Lyu2022} discovered that the selection of AGN using the X-ray-to-radio luminosity ratio was the most comprehensive of the multiwavelength selection methods they considered, until the recent multi-band JWST/MIRI measurements became available.  We also restrict our sample to those X-ray-bright \footnote{Where we refer to AGN in our sample as ``X-ray-bright" or ``X-ray-detected", we imply a source both detected in X-rays and confirmed to be an AGN using the criterion of a measurable and unmistakeable contribution from AGN emission processes in the corresponding X-ray portion of the SED. In other words, those sources showing an intrinsic X-ray luminosity higher than expected for stellar processes in a galaxy ($L_{X,int} > 10^{42.5}$ ergs/s,  \citealp{Lyu2022}).} AGN with an available SED-derived $N_{H}$ measurement from \citet{Luo2017} or \citet{Liu2017}. The vast majority of the resulting sample are optically faint, which minimizes AGN contamination in the rest-optical NIRCam imagery we utilize for morphology analysis.

In Figure~\ref{fig: full_sample_properties_hist} we show the distribution of the SED-derived AGN and host galaxy properties of the 243 X-ray AGN covered by the JADES/NIRCam GOODS-S and GOODS-N mosaic images utilized for morphology measurements. Where SEDs were not complete, the bolometric luminosity ($L_{bol}$) values shown represent the K-corrected intrinsic hard X-ray luminosity \citep{Lyu2022,Duras2020}; and the stellar masses ($M_{*}$) derived from \textit{g-i} colors and \textit{i}-band luminosity \citep{Lyu2022, Zibetti2009}.

\subsubsection{JWST/MIRI Mid-Infrared AGN Subsample}
\label{sec: sample_data}

The mid-IR AGN included in our primary sample are the X-ray-faint (i.e., X-ray undetected) subset of the newly identified, JWST/MIRI-selected AGN from SMILES, which covers $\sim$34 arcmin$^2$ \citep{Lyu2024a}. These sources were classified as AGN through an SED analysis consisting of twenty-seven photometric bands, including five HST/ACS bands (0.44 - 0.9 $\mu$m), fourteen JWST/NIRCam bands (0.9 - 4.4 $\mu$m), and eight JWST/MIRI bands (5.6 - 25.5 $\mu$m) \citep{Lyu2024a}. As we require the selected subset of (intrinsically luminous) mid-IR sources to lack a confident identification as X-ray AGN using the selection methods discussed in \citet{Lyu2022} and \citet{Lyu2024a}, these AGN are expected to represent a direct extension of the X-ray-bright subsample into the most heavily dust- and gas-obscured, Compton-thick (CT) regime. Their $N_H$ values were simulated via random draws from a normal distribution centered on a mean value of $10^{24.5}$ $cm^{-2}$, with the same standard deviation characterizing the distributions of real $N_H$ values in the X-ray `moderate' and `high' $N_H$ bins (see Figure ~\ref{fig: vis_stat_bar_plots}).

Of the total of 182 mid-IR AGN meeting the criteria above, 138 were covered by the JADES/NIRCam mosaics utilized in our analysis. The AGN and host galaxy properties are shown in Figure~\ref{fig: full_sample_properties_hist} alongside those of the X-ray sample, which were derived using the same SED fitting methodology as in \citet{Lyu2022} but with a number of refinements to the SED templates, as detailed in \citet{Lyu2024a}. Their $L_{bol}$ and $M_{*}$ values are based on their complete, broadband SED fits \citep{Lyu2024a}.

It can be seen that the mid-IR AGN exhibit higher median values of $L_{bol}$ and \textit{z} and lower median $M_{*}$ values than the X-ray AGN, the potential significance of which is discussed in Section~\ref{sec: discussion}. In order to enable a valid comparison between the two subsamples such that they together comprise a uniform sample, we conduct our primary analysis on the subset matched on $L_{bol}$, $M_{*}$, \textit{z}, as described in Section \ref{sec: design}.

\subsection{Comparison Low-Redshift Sample from the Swift/BAT All-Sky Survey}

Given the apparent incompleteness of our primary sample at $z<1$ on account of the limited combined sky area coverage of the GOODS-N and GOODS-S fields, we also include a comparison sample of local, Type I and II, hard-X-ray-selected AGN derived from the all-sky \textit{Swift} BAT Survey of AGN. This survey observed 74\% of the sky at northern Galactic latitudes above a limiting flux of $0.5 \times 10^{-10}$ ergs $cm^{-2}$ $s^{-1}$ in the 14-195 keV band \citep{Tueller2008}.


The complete sample of local X-ray AGN in \citet{Tueller2008} was identified in \textit{Swift} XRT follow-up observations of both previously known and new AGN candidates from the initial \textit{Swift} BAT all-sky survey. For our study, we selected the $83/103$ of these AGN for which there is available an $N_H$ measurement and a photometric flux appropriate for estimating stellar mass (\textit{J} magnitude), as well as optical imagery for measuring their host galaxy morphologies. 

We utilized the Panoramic Survey Telescope and Rapid Response System (Pan-STARRS1, or PS1) image cutout service \citep{Chambers2016} to obtain \textit{g, r, i, z, y} optical images of the local AGN comparison sample to measure their host galaxy morphologies. The 1.8-meter PS1 telescope uses the Gigapixel Camera 1 (GPC1) with a 7 $deg^{2}$ field-of-view and pixel scale of 0.258''/pixel. While the pixel scale of GPC1 is approximately 10 times larger than that of NIRCam, the much closer proximity of the $z<0.1$ comparison AGN sample than the primary sample at $z>0.6$ effectively cancels this loss in image resolution. For example, in a Universe obeying a flat cosmology with parameters $H_o$ = 69.6, $Omega_M$ = 0.286, and $Omega_{vac}$ = 0.714, each arcsecond subtended by a galaxy at $z=0.05$ corresponds to 0.984 kpc, while at $z=1.5$ an arcsecond corresponds to a physical size approximately 10 times larger, at 8.602 kpc. Therefore, the morphology measurements using the PS1 images for the local AGN sample may be directly compared to those derived from the NIRCam imagery containing the primary AGN sample.

In Figure~\ref{fig: full_sample_properties_hist} we show the bolometric luminosities ($L_{bol}$) and stellar masses ($M_{*}$) we calculated for this sample, which we use in our analysis (see Section ~\ref{design}) to match to galaxies in our primary sample. We calculated $L_{bol}$ by converting the available intrinsic 14-195 keV X-ray luminosity values to the 2-7 keV X-ray luminosity using Equation 1 in \citet{Rigby2009}; then applied the hard X-ray bolometric K-correction derived in \citet{Duras2020}. For the $M_{*}$ estimates, we used the available 2MASS \textit{J}-band fluxes as a proxy for the \textit{K}-band, given that the SEDs of low-redshift, hard-X-ray-selected AGN have $J/K\sim1$; and because the 2MASS survey is more sensitive in the \textit{J}-band than \textit{K}-band \citep{Tueller2008}. Finally, we applied the appropriate mass-to-light ratio ($M/L_{Ks}$) from \citet{Bell2009}.

\subsection{JWST NIRCam Data}
\label{sec: nircam}

\begin{figure}[h!]
    \centering
    \includegraphics[width=0.7\linewidth]{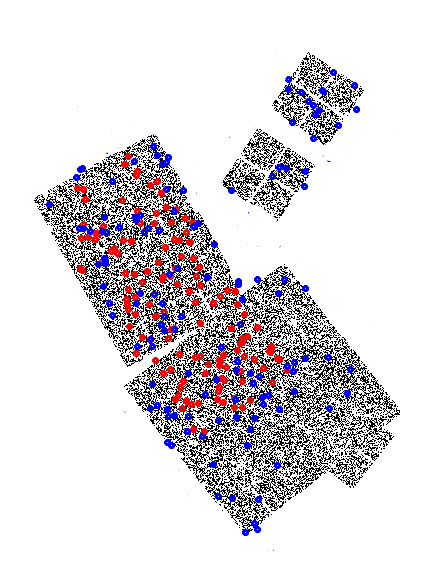}
     \centering
    \includegraphics[width=0.7\linewidth]{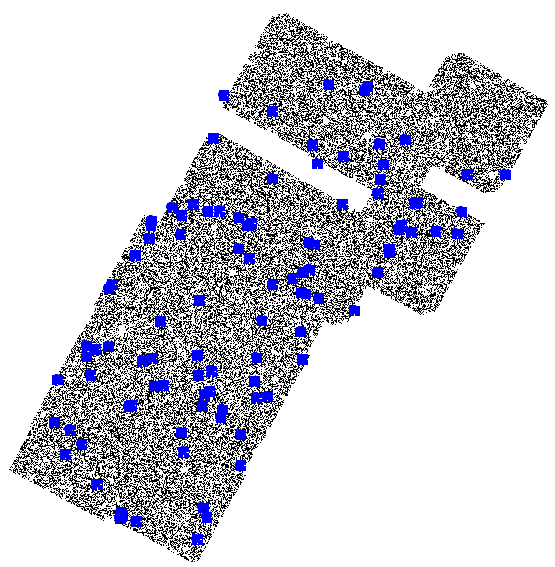}
    \caption{JADES version 1 NIRCam F150W GOODS-S (top) and GOODS-N (bottom) mosaic images containing the AGN sample. Blue points mark a subset of the X-ray-identified AGN from \citet{Lyu2022} and \citet{Lyu2024b}, and the red points the mid-IR-selected MIRI AGN from \citet{Lyu2024a}.}
\label{fig: mosaics}
\end{figure}

\begin{figure*}[h!]
    \centering
    \includegraphics[width=0.85\linewidth]{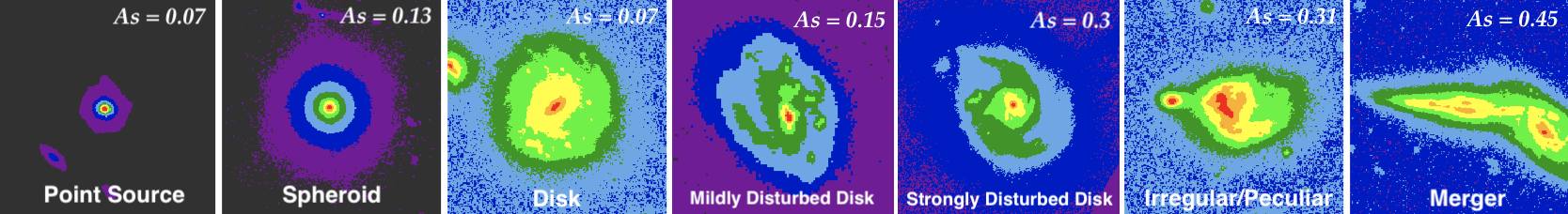} 
    \includegraphics[width=0.85\linewidth]{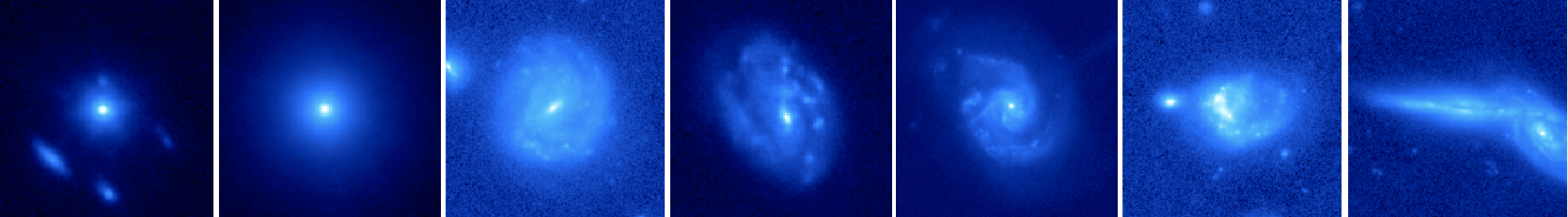}
    \includegraphics[width=0.85\linewidth]{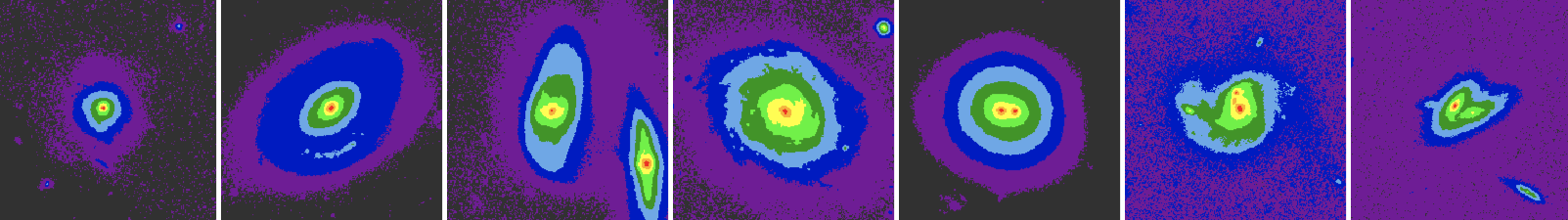}
    \includegraphics[width=0.85\linewidth]{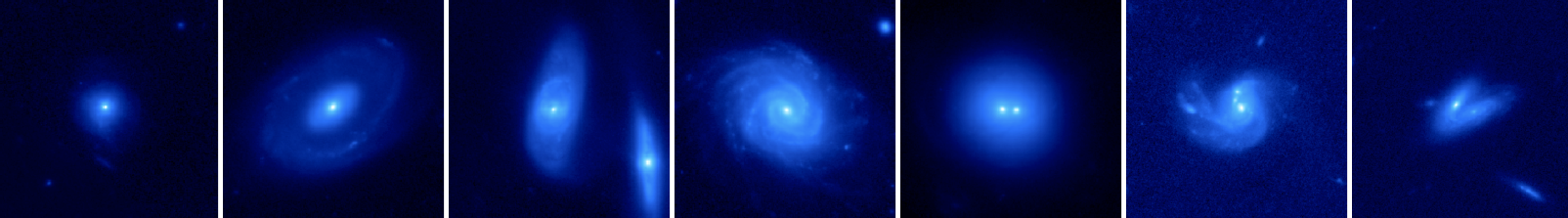}
    \includegraphics[width=0.85\linewidth]{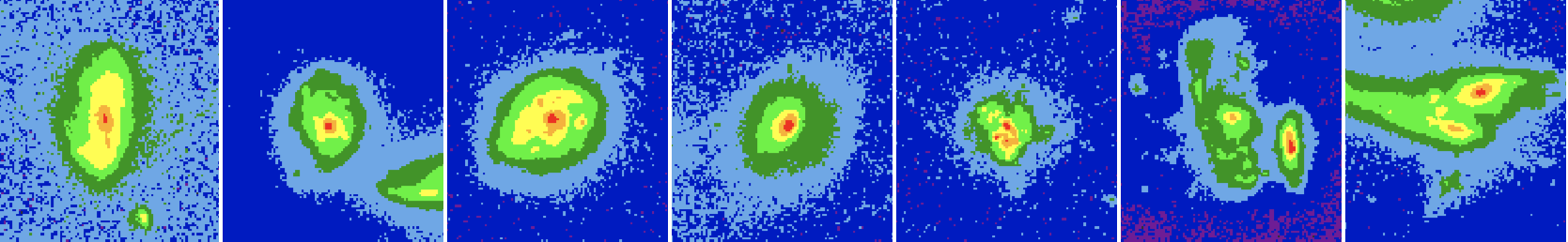}
    \includegraphics[width=0.85\linewidth]{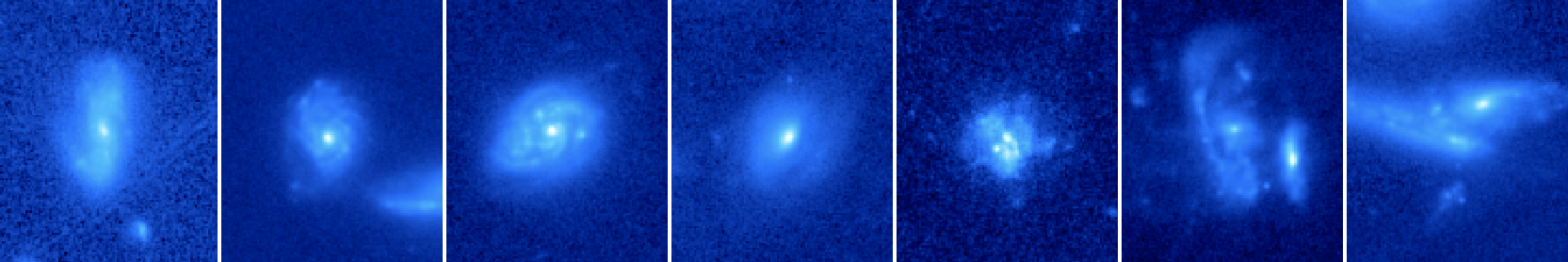}
    \includegraphics[width=0.85\linewidth]{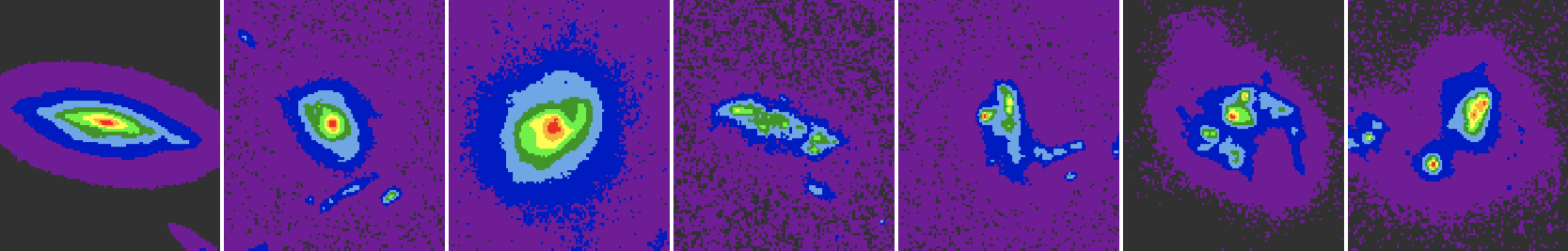}
    \includegraphics[width=0.85\linewidth]{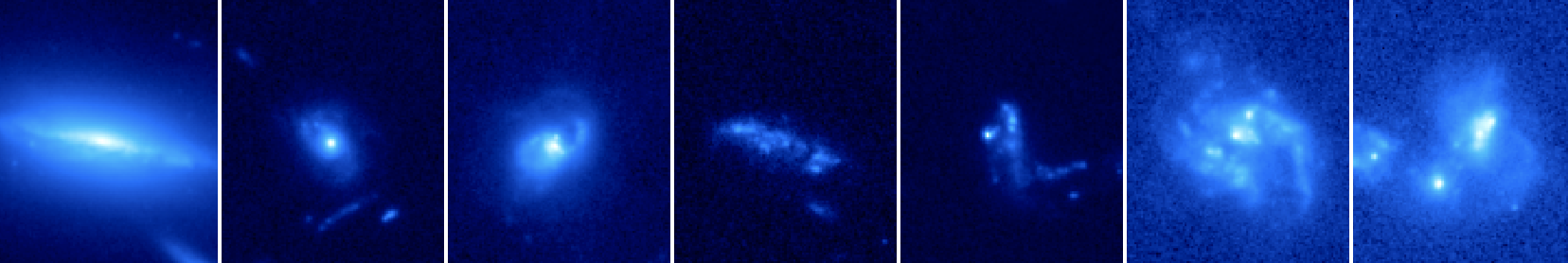}
    \caption{Examples of $0<z<1$ JADES v1 NIRCam F150W mosaic cutout images used to characterize rest-optical AGN host galaxy emission, with the full variety of morphologies observed displayed in the top panel (noting the non-uniform cutout sizes of all images shown, chosen in each case to highlight the details of the spatial morphology). Also displayed in the top panel is the shape asymmetry ($A_{S}$) measure corresponding to the qualitative morphology classification resulting from the visual analysis of the examples chosen. The four sets of galaxy images shown, from top to bottom, correspond to the following $N_H$ bins: $log(N_H)<22$, $22 \leq log(N_H)<23$, $23 \leq log(N_H)<24$, $log(N_H)\leq 24$. It can be seen that the galaxy morphologies range from visually undisturbed/symmetric to strongly disturbed/asymmetric shapes, including types that appear as both disturbed and disordered, i.e. with no discernable shape (noting that $A_S$ does not measure pixel brightness asymmetries like the classic asymmetry parameter, \textit{A}, but only spatial asymmetries in the binary detection mask associated with the galaxy emission; see Section ~\ref{sec: computer_vision} and Figure ~\ref{fig: statmorph} for details and examples). For the visual analysis task, the images were examined using a variety of SAOImage DS9 scales, stretches, colors, and zoom levels to ensure the most accurate morphological classification, i.e. to reveal the full spatial extent of the galaxy emission, as demonstrated here with two example colors (`cool' and `aips0') per set of sources.} 
\label{fig: images}
\end{figure*}

The NIRCam imaging utilized in this study was obtained as part of the JADES survey in JWST program 1180 (PI: Eisenstein), described in full detail in the JADES Data Release v1 paper \citep{RiekeM2023}, as well as the JWST Extragalactic Medium-band Survey (JEMS) (JWST PID 1963) \citep{Williams2023}. JADES v1 NIRCam mosaic images of the GOODS-S/HUDF and GOODS-N fields are available in the $0.9-4.4$ $\mu$m range in each of the following wide and medium filters: F090W, F115W, F150W, F182M, F200W, F210M, F277W, F335M, F356W, F410M, F430M, F444W, F460M, F480M. Each mosaic image achieves $0.03''$/pixel resolution and an AB magnitude depth of 29-30 \citep{RiekeM2023}.  We chose to work exclusively with the F150W NIRCam mosaic images for this study, as they proved to be the deepest and most resolved of the available mosaics that maximize coverage of the sky positions of the AGN in our sample (see the Appendix for a comparison of example GOODS-S mosaic image cutouts in all NIRCam filters available in the JADES v1 release). The sky projections of the NIRCam GOODS-S and GOODS-N mosaic images are displayed in Figure~\ref{fig: mosaics}, showing the overlapping coverage of the AGN sample; additional example galaxy image cutouts are shown in Figure~\ref{fig: images}.

Compared with the HST WFC3 F160W band that has been widely utilized in various types of galaxy studies, the NIRCam F150W band has a diffraction limit more than two times smaller and pixel sampling four times finer, and significantly greater sensitivity. The effective wavelength of 1.5 $\mu$m of this band samples the galaxy stellar emission in the rest-optical wavelength range of 0.3-0.9 $\mu$m for our $0.6<\textit{z}<3.8$ AGN sample. The choice to analyze the AGN host galaxy emission within this single observed waveband $-$ which probes a range of rest-frame wavelengths as opposed to a single consistent rest-frame wavelength $-$ was motivated by the following additional factors: 1) the limited overlap of our AGN sample with the mosaics exposed at the other NIRCam wavelengths sampled by JADES; 2) the resolution degradation of the diffraction-limited images at wavelengths longward of F150W\footnote{The observed degradation in image resolution for a given AGN host galaxy in our sample, as highlighted in the Appendix, when attempting to probe the same rest-frame wavelength by using mosaic images exposed at increasingly longer observed wavelengths, showed that no information was lost, or final conclusions altered, by foregoing their inclusion in our analysis. In other words, it was clear that by utilizing only the F150W-band mosaic, that we were always choosing the best imagery at hand to constrain galaxy morphology.}; and 3) the desire to avoid the extra sources of uncertainty that would be introduced to our analysis by comparing galaxy images extracted from mosaics in different filters.

\section{Methodology}

Methods to constrain galaxy morphologies historically range from manual, human visual classification to advanced, automated computational procedures, including machine learning where very large data samples and training data sets are available. Visual classification of galaxy shapes led to Edwin Hubble's famous `tuning fork' sequence of galaxy classes still in use today, as well as the equally famous and successful Galaxy Zoo project that employs citizen scientists around the world to visually classify galaxy images \citep{Lintott2008}. In the present study involving a humanly manageable data set, we utilize both human and computer visual classification methods to identify merger signatures in images of 464 AGN host galaxies (primary plus comparison AGN samples), in the form of a diagnosed galaxy-wide disturbance or asymmetry (see \citealp{Bignone2017} and references therein). 

The visual classification procedure chosen closely adheres to the one used in \citet{Kocevski2015} (which itself represents a refinement to the set of visual classification rules initially defined in \citealp{Kocevski2012}). We followed this procedure with a quantitative computer visual analysis in the form of an automated, non-parametric measure of the asymmetry in each galaxy's unweighted spatial pixel distribution using the \emph{statmorph} software \citep{RG2019}. 

We find several key advantages to employing a combination of these different types of methods in our characterization of AGN host galaxy morphology: 1) they provide an independent check of one another and therefore a reliable end result given the agreement of their respective findings; and 2) they complement one another in such a way as to refine the findings of the other, ensuring the certainty of a particular result (i.e., the binary classification of a galaxy as spatially disturbed or undisturbed, and presenting with or without a discernable shape, such as a disk or spheroid). A classical visual morphological analysis is robust against the inevitable sources of uncertainty introduced by an automated computer algorithm, where the latter is sufficient for constraining generalized, statistical properties, but sometimes fails to consider the nuanced features of individual galaxies that must be considered if one is to hope to achieve a complete understanding of the physical phenomena leading to their structures \citep{Reza2021, WY2022}. An example of this would be the distinctly human ability to readily distinguish between two galaxy images that were measured to have similar spatial and/or brightness asymmetries by a computer algorithm, but that clearly appear as belonging to distinct galaxy morphological classes. On the other side, the computer vision analysis lacks the inevitable observational bias imposed by the human eye, and is more sensitive to faint image features that could go undetected by a human.

\subsection{Visual Classification}

\citet{Kocevski2015} present a straightforward procedure for classifying the optical appearance of a galaxy into the following categories: undisturbed disk, disturbed disk, spheroid, point source, irregular, or galaxy merger. To streamline the results of our analysis, we refined the `disturbed disk' category by splitting it into two sub-categories: ``strongly" and ``mildly" disturbed, as we (qualitatively) observed significant numbers of both types in our trials. Furthermore, we classify as ``disturbed'' all galaxies appearing as irregular or peculiar (i.e., disordered, lacking a clear, discernible shape); an obvious merger, such as a close pair with two visible nuclei, or a `trainwreck' appearance; as well as an asymmetric disk. Likewise, all point sources, spheroids, and apparently symmetric disks were classified as ``undisturbed." Finally, we classified the small number of apparently fully edge-on disk galaxies as ``undisturbed", under the presumption that a galaxy-wide disturbance would manifest itself in three dimensions and therefore be observable from all disk inclination angles relative to the observer. Representative examples of the seven morphology classes we adopted for the visual classification task are displayed in the top row of Figure ~\ref{fig: images}.

\subsection{Computer Vision}
\label{sec: computer_vision}
While visual classification tasks performed by a human have proven to be very accurate for many types of analyses \citep{Brinchmann1998, Bundy2005, Kampczyk2007, Darg2010, Treister2012, Kartaltepe2015, WY2022}, by definition they are subject to `human error' on account of oftentimes uncontrollable factors such as imperfect vision, and bias and differences in reasoning. Another important consideration is the nonlinear response of the human eye to light intensity, whereby weaker disturbances could go unnoticed in cases of high brightness contrast, or modest image resolution or signal-to-noise, that would lead to mis-classification of truly disturbed galaxies as symmetric/undisturbed. Therefore, to systematically quantify the results of our visual analysis and catch such cases of subtle host disturbances that may have gone unnoticed or mis-categorized in the visual classification task, we utilized the \emph{statmorph} software package to perform an independent, computer visual check of the results. This software package was specifically chosen for its proven ability to detect faint merger signatures around the edges of galaxies via it calculation of the \textit{shape asymmetry} parameter, described in the next section \citep{RG2019}.

\subsubsection{Metric for Disturbance Level}

The use of computational methods to characterize galaxy morphologies allows for a range of metrics to determine the level of disturbance.  The commonly used suite of non-parametric galaxy morphology indicators includes \textit{asymmetry} (\textit{A}), \textit{clumpiness} (\textit{S}), the \textit{concentration index} (\textit{C}), the Gini index (\textit{G}), and the \textit{moment of light} (\textit{$M_{20}$}) (\citealt{Lotz2004}, \citealt{Wu1999}, \citealt{Bershady2000}, \citealt{Conselice2000}, \citealt{Schade1995}, \citealt{Abraham1994}, \citealt{Abraham1996}, \citealt{Iss1986}, \citealt{Takamiya1999}, \citealt{Conselice2003}). A relatively recent addition to this set of morphology indicators is the \textit{shape asymmetry} ($A_S$), introduced in \citet{Pawlik2016} to explicitly automate the recognition of galaxies with faint asymmetric tidal features suggestive of an ongoing or past merger. \citet{Pawlik2016} showed that $A_S$ far outperforms the classic \textit{A} parameter in detecting low surface brightness features around the edges of a galaxy; and the merger simulations of \citet{Nevin2019} show it to be the most accurate single morphology indicator at identifying merger morphologies, only marginally surpassed by a measure that combines $A_S$ with \textit{C}, \textit{G}, and \textit{A}.

$A_S$ is calculated in the same way as \textit{A}, except the measurement is performed on the binary detection mask (segmentation map) corresponding to an image of galaxy emission, rather than the galaxy image itself (see \citealp{Pawlik2016} and \citealp{RG2019} for further details). Therefore, while \emph{A} considers asymmetries in both the spatial and brightness distributions of the pixel values in a galaxy image, $A_S$ considers only those due to a spatial asymmetry, thereby isolating disturbances that can be attributed to a physically disruptive event such as a major merger, as opposed to an internal galaxy sub-structure unrelated to merging activity that would qualify as a brightness asymmetry.\footnote{This feature also minimizes the impact on our morphology measurements from the effect of dust attenuation of starlight, which has been shown to bias measures of galaxy sizes and morphologies (see, e.g., \citealp{Suess2022}). Given that $A_S$ effectively traces the outline of a galaxy's shape in a binary image, without regard to differences in pixel intensity, dust attenuation would only bias an $A_S$ result for a given galaxy if its outer edges were significantly dusty, and the starlight significantly absorbed, fully or partially hiding the full extent and shape of the host galaxy.}

Another way in which $A_S$ outperforms the other non-parametric galaxy morphology indicators at identifying merger signatures is its ability to maximize the \textit{timescale} over which both major and minor mergers may be identified.  \citet{Lotz2008} and \citet{Bignone2017} correlate key non-parametric morphology measures with the recent merger history of galaxies to show that the $Gini-M_{20}$ criterion and the classic asymmetry (\textit{A}) parameter are effective at identifying both major and minor mergers, but only completely and effectively in the pre-coalescence phase while a double nucleus is visible. Sampling well the pre- and post-merger phases is very important in the context of our study given that only a small handful of our entire sample meet the double-nucleus criterion that is often the only merger stage considered in merger studies. Furthermore, inspection of the environs of each AGN within 50 kpc for a potential merging companion at the same redshift turned up only a small handful of pre-coalescent merger candidates. This supports that the window of time during which active merging occurs is small in comparison to the pre-and post-merger phases. Furthermore, given the significantly longer timescale of observability ($\sim2$ Gyr) of the remnants of major mergers as compared to minor \citep{Lotz2010, Bignone2017}, the number of isolated disturbed hosts identified with $A_S$ offers an additional measure of the distinction between the two in our sample. 

\begin{figure*}
\centering
  \includegraphics[width=1.0\linewidth]{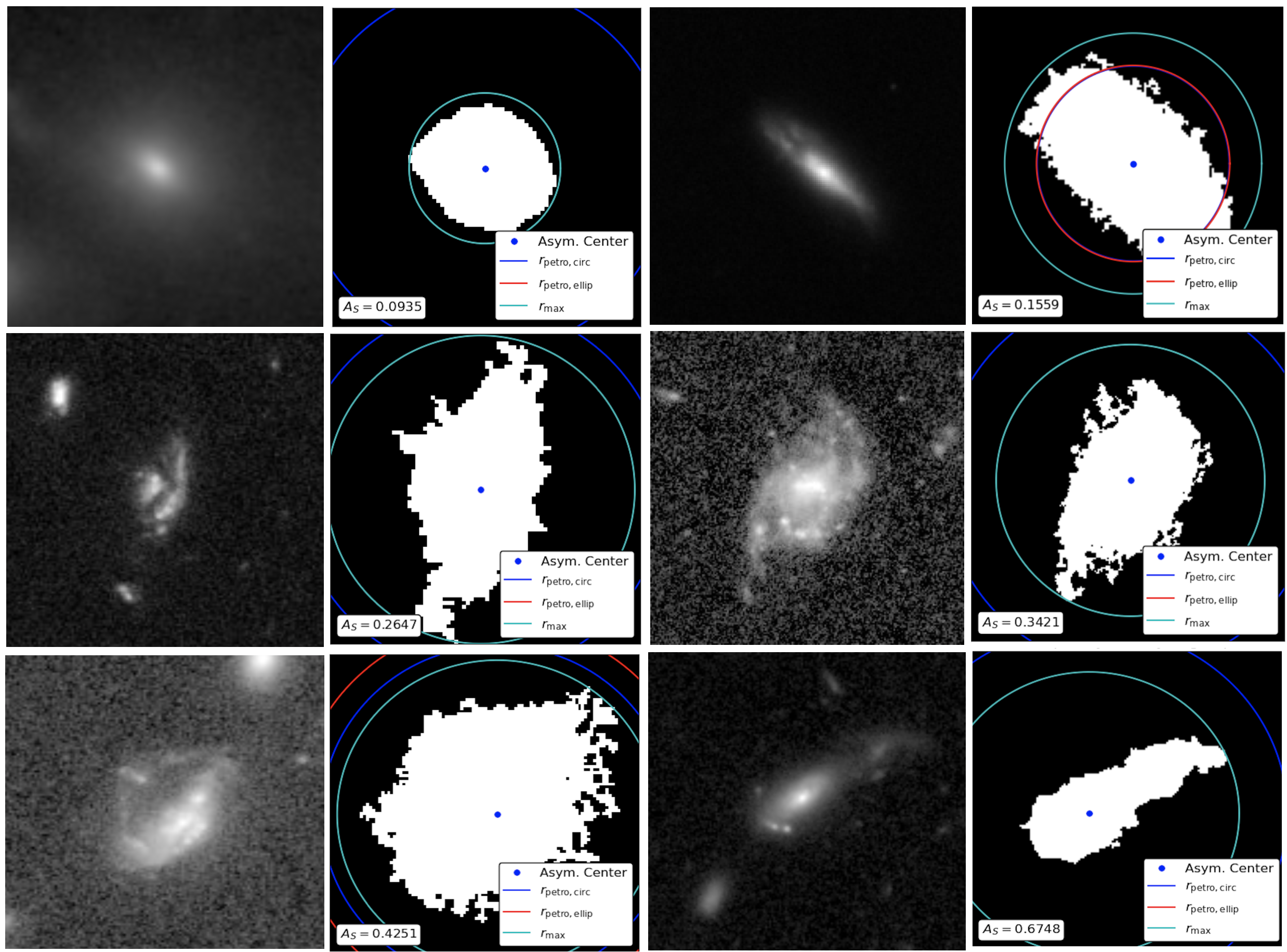}
  \caption{Examples of \textit{statmorph} figure output showing the binary detection mask (right image in each cutout image pair) that is used to calculate the shape asymmetry ($A_S$) of an input image of galaxy emission (left image in each cutout image pair, noting the difference in zoom level from the left to right image). $A_S>0.2$ marks a strong asymmetry/disturbance, values between 0.1 and 0.2 represent mild asymmetry, and $A_S<0.1$ is considered symmetric/undisturbed; the measurement is performed within the circular aperture set by $r_{max}$ (cyan line). The \textit{statmorph} output image also displays the circular and elliptical Petrosian flux apertures that are used in various other morphology measures output by the code, where the circular aperture serves as the basis for the computation of the classic asymmetry (\textit{A}) parameter, as well as the sky background level that sets the brightness threshold for the binary detection mask (refer to \citet{RG2019} for further details). Note that the Petrosian apertures are not simultaneously visible in all figures generated by \textit{statmorph} due to differences in the input image cutout size, which we customized to each source to eliminate contamination from the source aperture.}
\label{fig: statmorph}
\end{figure*}

\begin{figure*}[h!] 
   \centering
     \includegraphics[width=1.\textwidth]{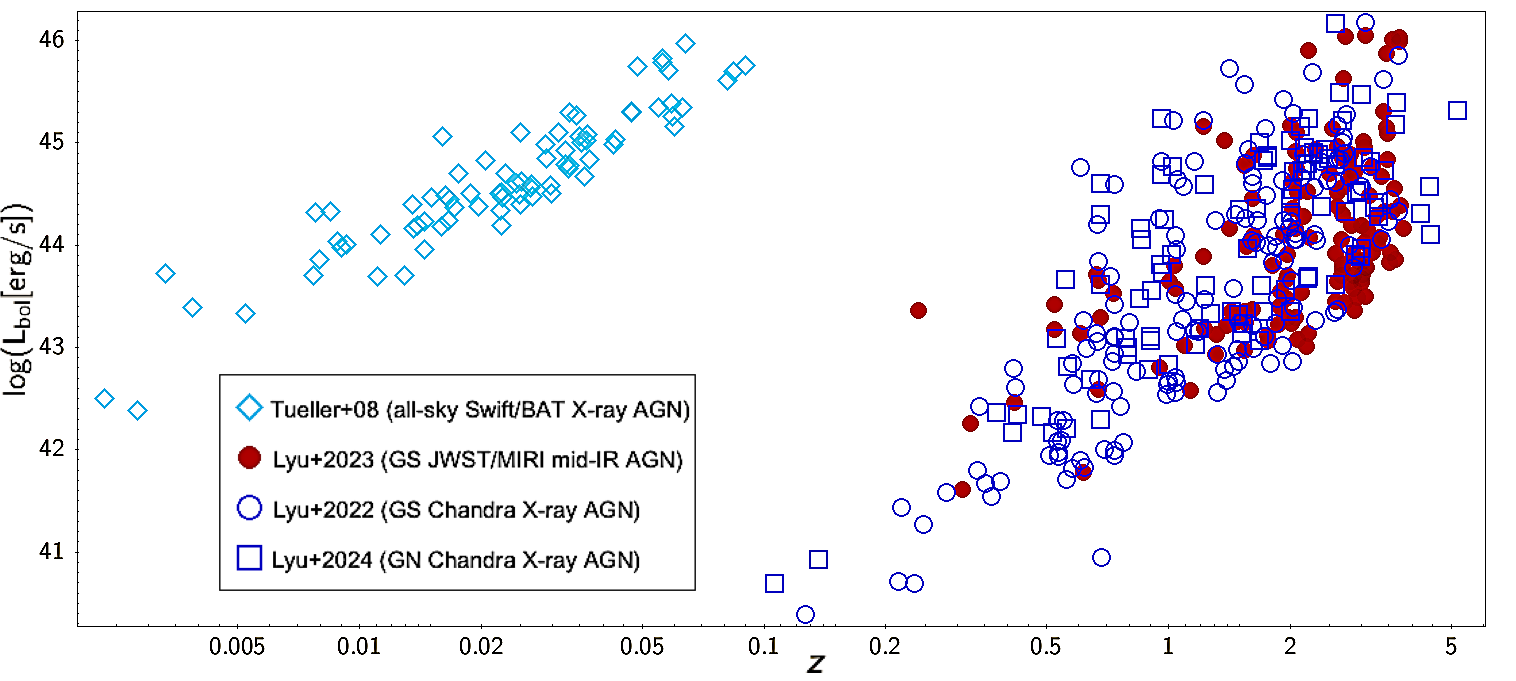}
        \centering
     \includegraphics[width=1.\linewidth]{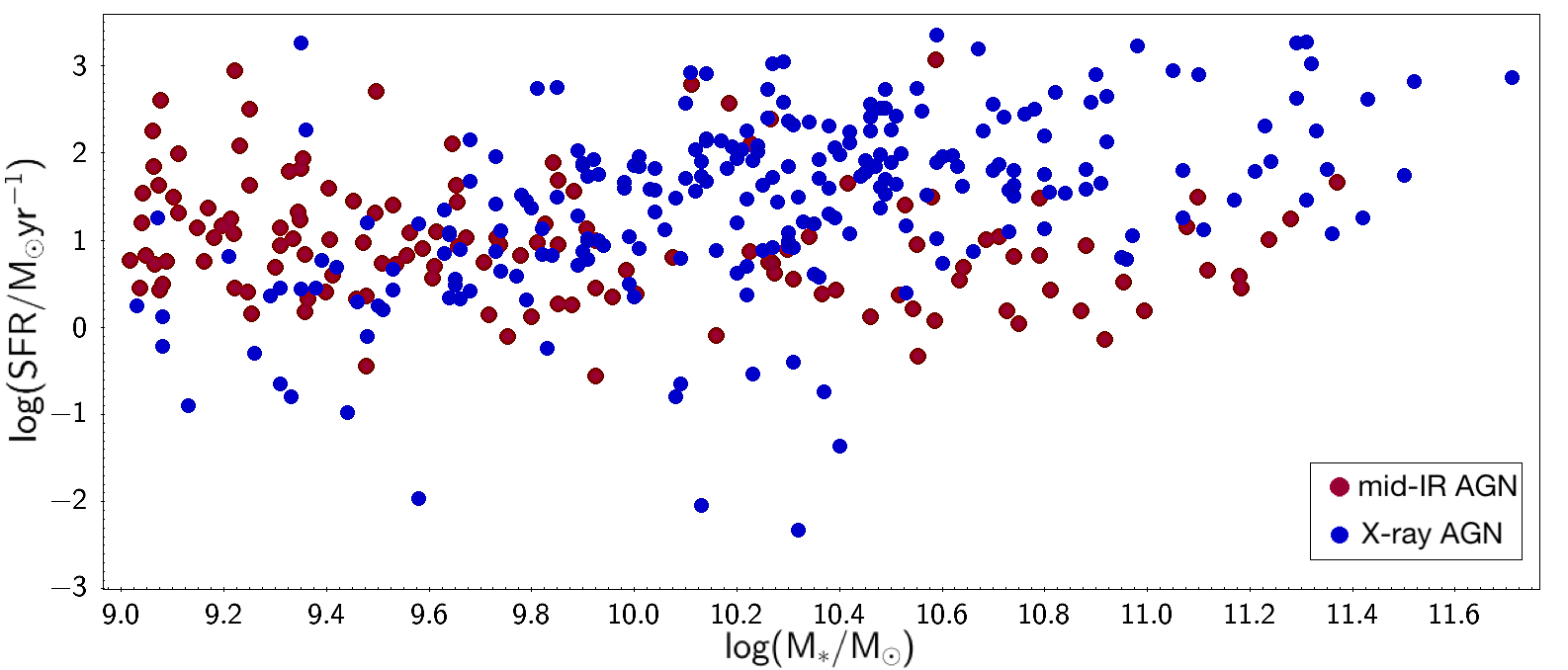}
     
    \caption{Bolometric luminosity and redshift comparison of the distinct components of the primary sample drawn from \citet{Lyu2022}, \citet{Lyu2024a}, and \citet{Lyu2024b}, and the low-redshift comparison sample from \cite{Tueller2008}. The incompleteness of our sample above the luminosity detection threshold at $0.1<z<0.6$ is visible, and is therefore excluded from the analyses discussed in this paper. Bottom: SED-derived (\citealp{Lyu2022,Lyu2024a,Lyu2024b}) star formation rates as a function of stellar mass for the primary sample, showing similar distributions and implied stellar populations for the X-ray- and mid-IR-selected subsamples.} 
\label{fig: sample_properties_cont}
\end{figure*}

The Illustris merger simulations of \citet{Bottrell2023} find that so-called mini mergers ($0.01 \leq \mu < 0.1$) are expected to occur significantly more frequently than minor ($0.1 < \mu < 0.25$) and major mergers ($\mu > 0.25$), and therefore may represent a significant, overlooked cause of galaxy morphological disturbance. While the simulations show that individual mini mergers are not likely to induce a significant observable disturbance, they reveal that the integrated (classic \textit{A}, light-weighted) asymmetry due to mini mergers over $0.01 \le z \le 0.7$ constitutes $70\%$ of the asymmetric structure in the mergers of their study. They therefore suggest that the aggregation of asymmetry enhancements by a series of mini merger events for a given galaxy may be driving the build-up of light asymmetry and lopsidedness in galaxies, as opposed to major or minor mergers. However, their study examined only TNG50 star-forming galaxies and adopted the classic asymmetry parameter as the measure of galaxy disturbance,  while we consider \textit{active} galaxies and utilize the \textit{shape} asymmetry parameter that ignores asymmetries in the galaxy surface brightness distribution. While the majority of our AGN sample are forming stars (see Figure ~\ref{fig: sample_properties_cont}), it is not clear if the results of this study are directly applicable to star-forming AGN. 

However, given that even minor mergers are generally considered too weak to produce significant morphological changes to the galaxies involved (e.g., \citealp{Martin2018}), in the following, we ignore mini mergers as a possible cause of the galaxy-scale spatial disturbances observed in our AGN sample. In Section ~\ref{sec: statmorph_results}, we report a preliminary estimate of the fraction of host galaxy disturbances likely to be caused by a major versus minor mergers, based on the merger simulations analyzed in Nevin et al. (2019), and perform a more extensive analysis in Paper II. 

\subsubsection{Implementation of {\it statmorph}}

The \emph{statmorph} computer code reads a galaxy image and its associated weight map (the $1\sigma$ error image, also known as the `sigma image' in \emph{Galfit} and similar image analyses) to examine both the brightness and spatial distribution of its pixel values. It then calculates numerous measures of galaxy morphology, including $A_S$ and the aforementioned suite of classic non-parametric morphology indicators (which are considered in combination with $A_S$ for an expanded and refined AGN host galaxy morphological analysis, in Paper II). 
The binary detection mask utilized in the \emph{statmorph} calculation of $A_S$ is defined as the contiguous group of pixels brighter than 1$\sigma$ above the mode of the pixel values in the background-subtracted, smoothed galaxy image, that includes the brightest pixel.  Examples of the \textit{statmorph} figure output showing the details of the $A_S$ calculation are displayed in Figure~\ref{fig: statmorph}. 
 
For each of the 464 galaxies analyzed in our combined primary and comparison AGN samples, we inspected and verified the automatically generated \textit{statmorph} results to ensure proper source detection and aperture size and location. We identified and remedied instances where obvious foreground objects or image features/artefacts obviously not physically associated with the source being analyzed, were being included in the aperture used to calculate $A_S$. This was done by iteratively redefining the source aperture until such features were no longer erroneously being included in the $A_S$ measurement. Therefore, the ultimate methodology we employed in our analysis of AGN host morphology was a synergistic union of human reasoning and computer vision.

\section{Experimental Design}
\label{design}

There is a history of different studies reaching different conclusions about the prevalence of mergers and disturbances in AGN hosts. In part, this range of results may arise from differing AGN samples, for example, from comparing those with significantly different AGN luminosities; or radio galaxies to radio-quiet ones; or AGN to insufficiently matched non-active control samples; as well as the comparison of merger studies with different adopted observational criteria for what constitutes the merger phase. To minimize the impact of such biases in our AGN sample, we ensure that the AGN subsets detected in the X-ray and the mid-IR are matched in physical properties as described in Section ~\ref{sec: design},  and therefore can be assumed to be drawn from the same AGN class and analyzed as a uniform, representative sample. Furthermore, given that differing analysis approaches may also lead to tension between different studies of AGN host morphologies, we calibrate our relatively new approach of classifying mergers with $A_S$ on the \textit{Swift}/BAT control sample to enable comparison with previous work. Figure~\ref{fig: sample_properties_cont} summarizes the main samples used in this study, showing their similar distributions of bolometric luminosity, star formation rate (and implicit host mass stellar populations), and stellar masses over the redshift range considered. 

\subsection{Calibration of methods on low-z samples}
\label{calibration} 

The \textit{statmorph} analysis of our local \textit{Swift}/BAT AGN comparison sample results in $33 \pm 6$\% with $A_S$ $>$ 0.2, indicating significant disturbance (where the error is statistical based on sample size). This value agrees well with studies of other samples, most notably the \textit{Swift}/BAT AGN sample analyzed in \citet{Kim2021}, which provides the most direct comparison to ours (our samples are both confined to $z<0.1$, and only differ in that we consider the subset with a measurement of $N_H$). They report that $18 - 25$\% of the host galaxies in their sample show features associated with mergers, e.g., tidal tails, shells, or major disturbances (our slightly higher value may be due to our incorporating computer vision classifications whereas \citealp{Kim2021} used only visual ones).

\citet{Kauffmann2003} find that about 30\% of a sample of 100 SDSS AGN at z $<$ 0.1 are disturbed or interacting. \citet{Kim2017} analyzed archival HST images of 235 AGN hosts at z $<$ 0.35, finding that 18\% of the broad line Type 1 hosts showed asymmetry parameters $>$ 0.2 indicative of strong disturbance. \citet{Zhao2019} show that  34\% of the host galaxies of 29 optically selected Type II quasars are mergers or interacting, and \citet{Zhao2021} find that $\lesssim$ 20\% of 35 low-redshift Palomar-Green (PG) quasars have hosts with evidence of major mergers. The incidence of disturbance may increase for the most luminous AGN \citep[e.g.,][]{Treister2012,Pierce2023, Comerford2024}; however, for AGN of moderate luminosity, it is clear that we can take $\sim$20 - 30\% as a well-determined and cross-calibrated incidence of significant host galaxy disturbance.  

\subsection{Approach to Measuring Morphological Disturbance vs. Obscuration}

In the following, we evaluate AGN host galaxy morphology as a function of AGN obscuration, characterized by $N_H$ (or our surrogate $N_H$ from infrared properties). To rigorously constrain a statistical obscuration-disturbance ($N_H$-$A_S$) correlation, we conduct our primary host galaxy morphology study on a subset of our primary sample with matching distributions of $L_{bol}$, $M_{star}$, and \textit{z}. We repeat this analysis on the full, unmatched primary sample plus local comparison sample, to investigate an apparent evolution of the $N_H$-$A_S$ correlation itself with redshift, taken in the context of the parallel evolution of the other average galaxy properties considered in our study.

Previous studies have compared AGN to matched samples of non-active field galaxies at redshifts similar to those in our study \citep[e.g.,][]{Kocevski2012}, finding no significant differences in the incidence of disturbances. However, constructing closely matched samples can be challenging  \citep[e.g.,][]{Gabor2009, Kocevski2012}, particularly at high redshift because the field galaxies are generally of lower mass than the AGN hosts. Furthermore, these studies, with the exception of \citet{Kocevski2015}, have not been extended to extremely obscured AGN; defining a suitable comparison sample for the heavily obscured AGN hosts is a future challenge.  In the present paper, we focus on a narrow scientific question that is independent of the behaviour of non-active galaxy counterparts, namely, to confirm if there is an increasing level of disturbance with increasing obscuration in the AGN of our sample.

\subsubsection{Controlling for Obscuration with a Propery-Matched Sample}
\label{sec: design} 
\begin{figure*}
   \centering
     \includegraphics[width=1\linewidth]{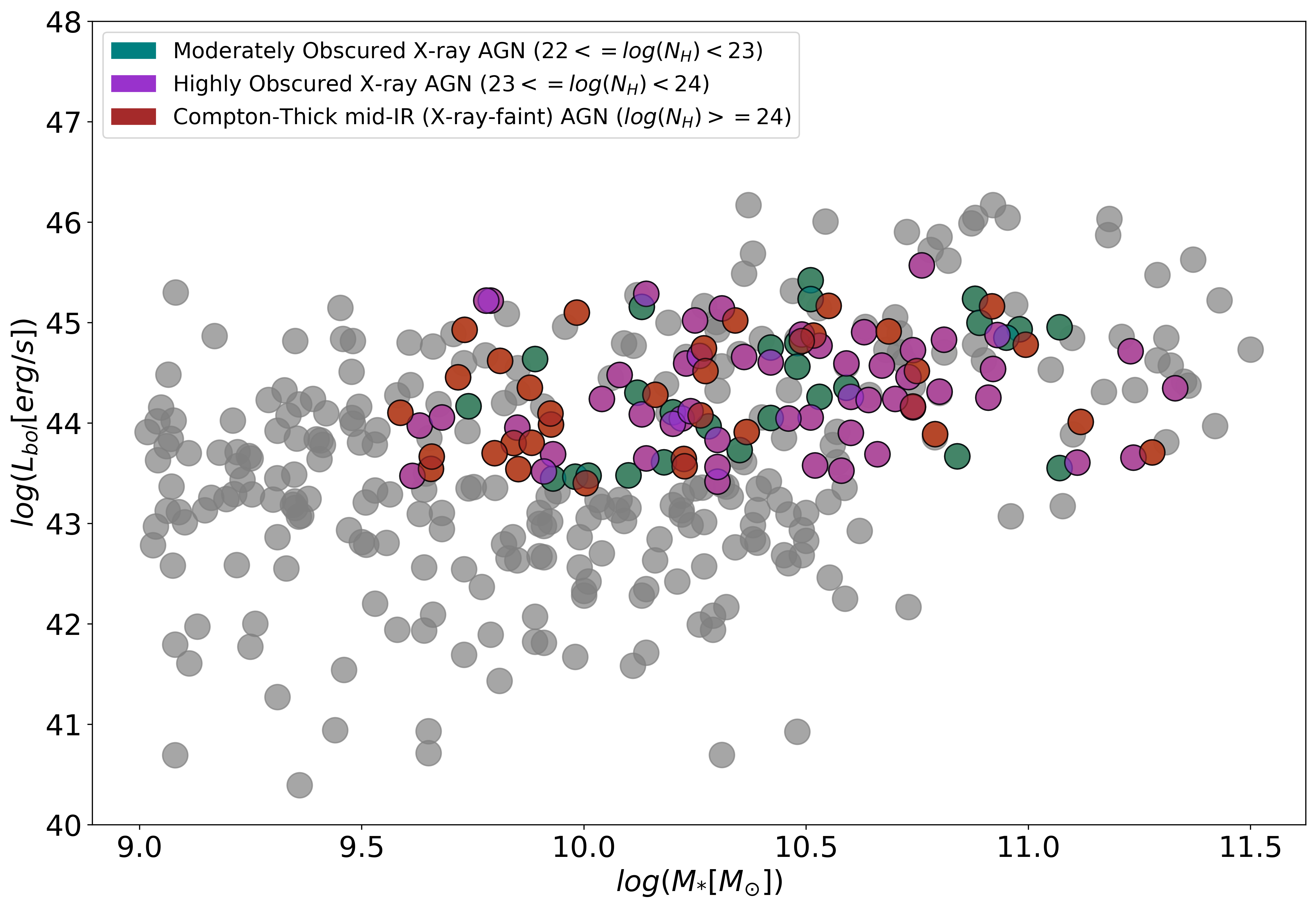}
    \caption{Top: Properties of the primary property-matched sample highlighted in colored symbols against the full, unmatched sample (gray symbols), including AGN with matching distributions of $M_*$, $L_{bol}$, \textit{z}, and $N_H$, used to constrain a correlation between host galaxy disturbance (as measured with $A_S$) and obscuration (as measured with $N_H$).} 
\label{fig: primary_properties}
\end{figure*}

In our analysis of AGN host galaxy morphology, we controlled for AGN obscuration across the sample to enable a measurement of a true correlation with morphological disturbance. This was carried out by first restricting our analysis to AGN in relatively massive host galaxies with $M_{*} > 10^9 M_{\odot}$, given the possibility that the dominant physical processes shaping the evolution of low-mass active galaxies may differ from those molding more massive ones; additionally, the sample is likely incomplete and potentially biased for lower-mass hosts \citep{Lyu2024a}. An additional cut on the sample was made to isolate only those AGN host galaxies with matching distributions of $N_H$, $M_*$, $L_{bol}$, and \textit{z}, as shown in Figure~\ref{fig: primary_properties} (in Figure~\ref{fig: sequence}, the mean and standard deviation of each property of the matched sample is plotted in comparison to the corresponding values of the full sample).  This resulted in a statistically significant matched sample (113 sources, 30\% of the primary sample) of moderately to heavily obscured AGN host galaxies bound by the following parameter space: $9.5<log(M_{*})\le11.4$ $M_{sun}$, $43.4<log(L_{bol})\le45.5$ ergs $s^{-1}$, $N_H>=22.0$, and $0.6<z<2.4$. This process naturally sifted out those portions of the sample influenced by a number of key observational biases, namely 1) redshift bias from the physical evolution of $L_{bol}$ and $N_H$ with redshift (refer to Figures 10 and 17 and associated discussion in \citealp{Lyu2022}); 2) bias due to sample incompleteness on account of limited sky area coverage at low redshift; and 3) selection bias against intrinsically faint sources at high redshift (i.e., the AGN luminosity range of our matched sample lies above the unobscured X-ray detection limit at all redshifts sampled). Furthermore, the restricted redshift range of the primary matched sample mitigates the effect of a bias that would be introduced by mixing sources lying at vastly different distances, due to the unequal probability of detecting morphological features in such a case due to differing surface brightness limits (see, e.g., \citealp{Bamford2009}). 

Finally, we split the remaining sample into bins of low ($N_{H}<10^{22}$ $cm^{-2}$), moderate ($10^{22} \le N_{H}<10^{23}$ $cm^{-2}$), high ($10^{23} \le N_{H}<10^{24}$  $cm^{-2}$), and CT ($N_{H} \ge 10^{24}$ $cm^{-2}$) levels of obscuration, using the $N_H$ hydrogen absorbing column density parameter as the measure of obscuration. The low-, moderate-, and high-$N_H$ bins consist of the X-ray-selected subsample, described in Section 2.1.2, with their $N_H$ values derived from spectral fitting (\citealp{Luo2017, Liu2017}). The CT bin includes the mid-IR-selected AGN, which are presumed to represent a direct extension of the X-ray-bright sample into the X-ray-faint CT regime (whose $N_H$ values were simulated as described in Section ~\ref{sec: sample_data}.

\subsubsection{Contextualizing the Obscuration-Disturbance Trends in the Full Sample}

\begin{figure}[h!]
    \centering  
         \centering
         \includegraphics[width=1\linewidth]{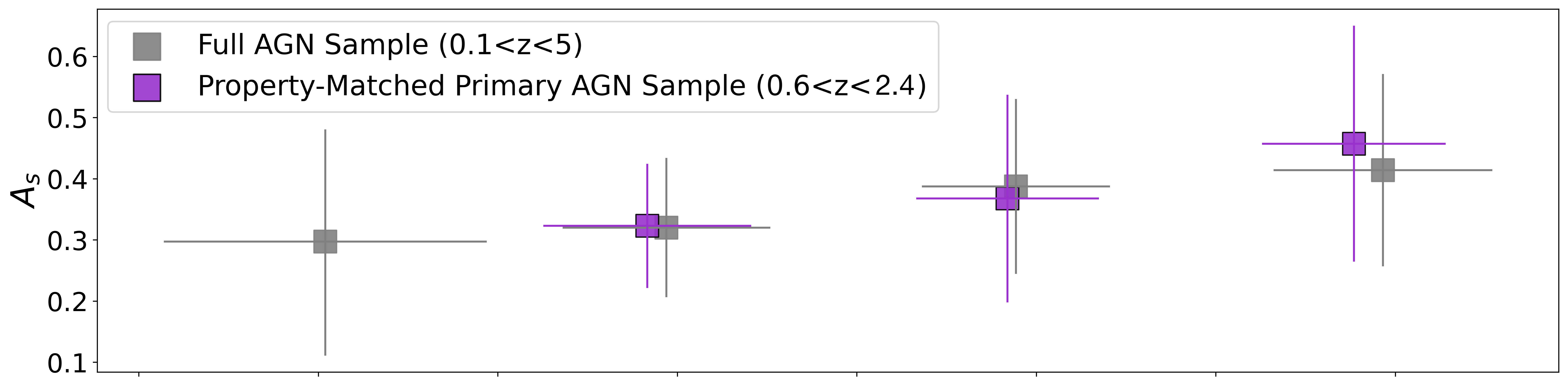}
         \centering
         \includegraphics[width=1\linewidth]{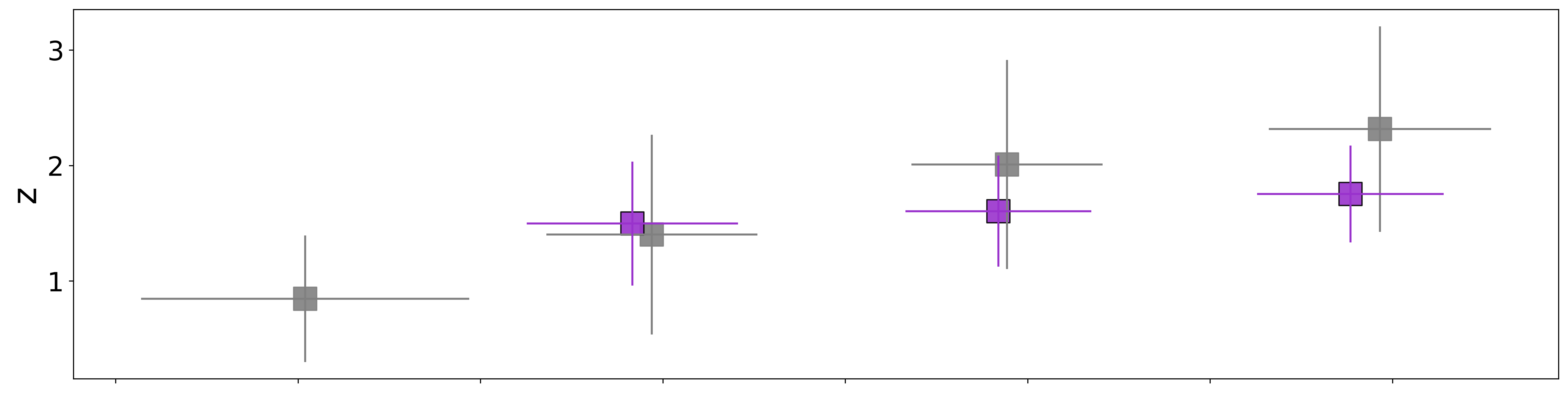}
         \centering
         \includegraphics[width=1\linewidth]{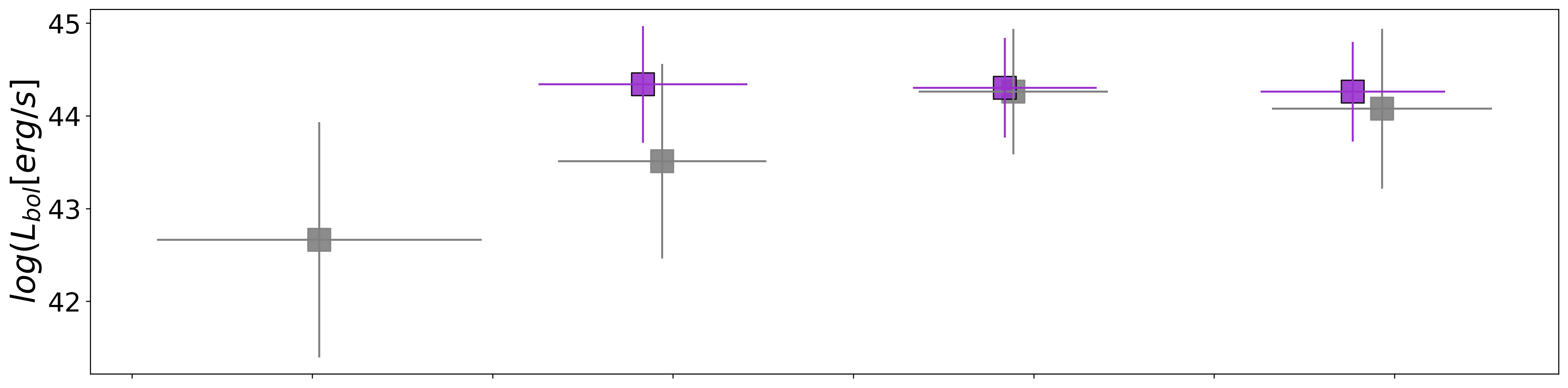}
         \centering
         \includegraphics[width=1\linewidth]{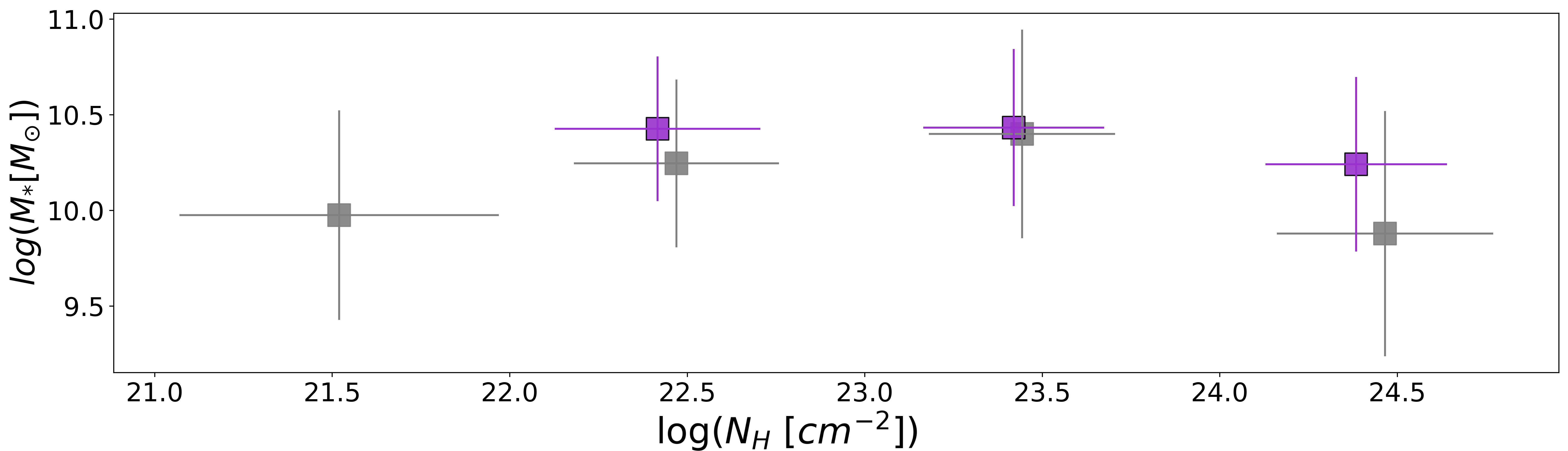}
         \caption{Average AGN host galaxy properties per bin of obscuration ($N_H$) considered in the investigation of an obscuration-merger correlation in the property-matched primary sample (purple points), with the corresponding values for the full, unmatched primary sample shown in gray symbols, for context (including the incomplete data bin at $0.1<z<0.6$. The error bars indicate the sample standard deviation. (A similar plot is shown in Figure \ref{fig: finalfit} but including the local $z<0.1$ comparison AGN sample in addition to the primary AGN sample).}
\label{fig: sequence}         
\end{figure}

To explore the apparent evolution of the $N_H-A_S$ correlation as a function of redshift within our AGN sample, and the implications of such a difference in AGN character at $z<0.1$ in comparison to $z>\sim0.6$, we performed an additional analysis trial utilizing the full, unmatched primary sample in conjunction with the local comparison sample ($0<z<0.1$). We divided the AGN into the redshift bins that naturally arose in the galaxy parameter space, grouping the sources with complete and uniform distributions of $L_{bol}$ (i.e., following the evolving AGN detection threshold in $L_{bol}$ versus \textit{z} space): $z=0-0.1$ (spanning 1.3 Gyr), $z=0.6-1.4$ (3.4 Gyr), $z=1.4-2.4$ (1.8), $2.4-3.2$ (0.7 Gyr), and $z=3.2-3.8$ (0.4 Gyr), noting that we excluded sources at $z=0.1-0.6$ due to sample incompleteness in this redshift range. While the source properties in these redshift bins were not matched, they do have similar average values of $M_{*}$, as displayed in Figure~\ref{fig: finalfit} and Figure~\ref{fig: sequence}.  This trial allowed for the inclusion of the lowest-obscuration/lowest-redshift AGN in the $N_H$-$A_S$ correlation investigation, given that they failed to pass the full set of property-matching criteria for the primary matched sample. These sources constitute a minority of the sample due to an apparent true deficiency of unobscured AGN at higher redshifts; and possibly also due to our sample incompleteness at $0.1<z<0.6$.

\section{Results}
\label{sec:results}

\subsection{Overview}

The results of both the human and computer morphology characterization procedures largely corroborate one another, and reveal two unexpected features of the Seyfert population examined: 1) the vast majority of the full sample exhibit the telltale disturbed morphological signatures of merging activity; and 2) the dominant morphological type characterizing the X-ray-detected versus mid-IR-bright/X-ray-faint subsets differ.

\begin{figure*}
   \centering
     \includegraphics[width=1\linewidth]{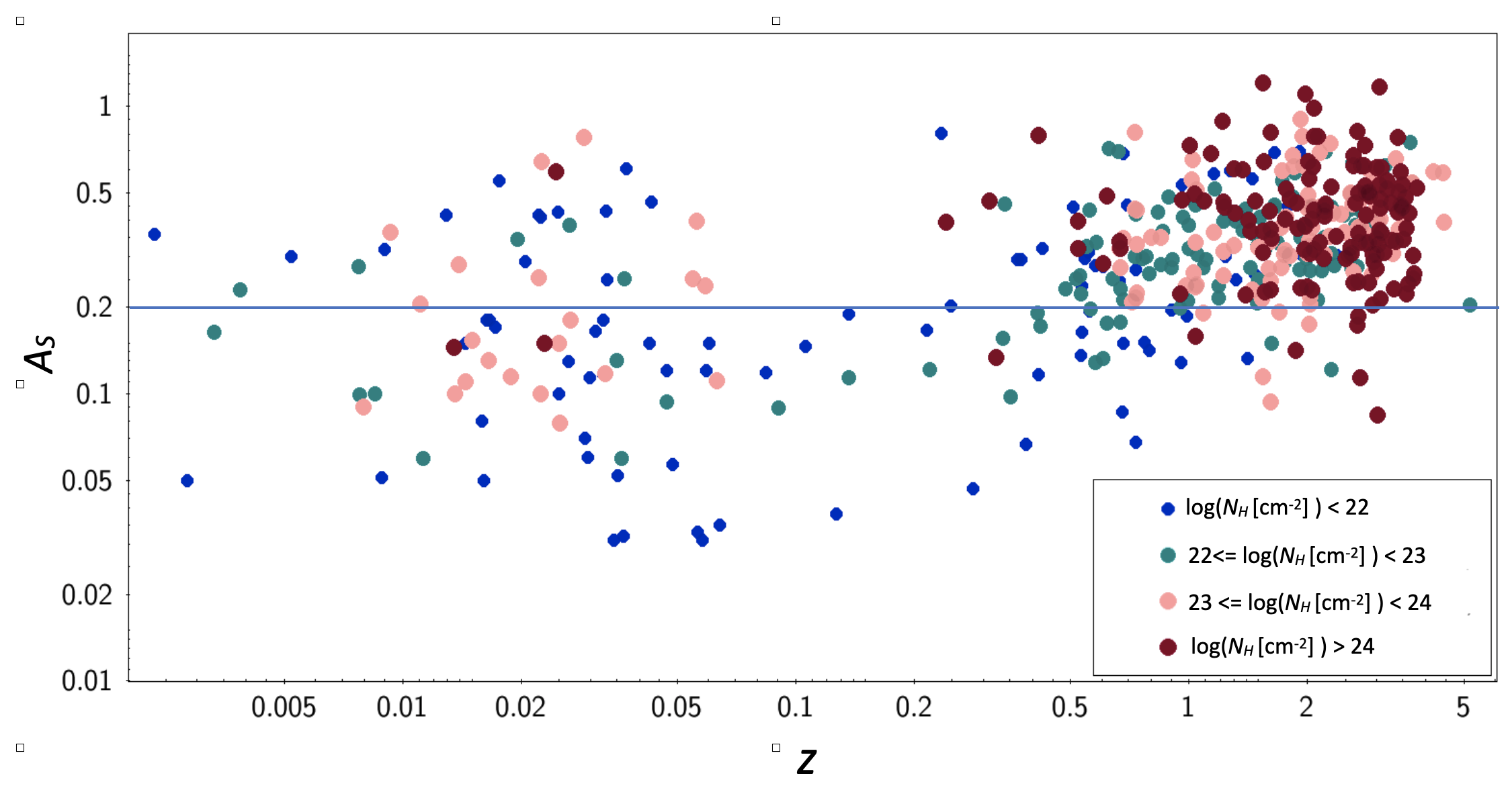}
     \centering
    \caption{The distribution of the disturbance ($A_S$) and obscuration ($N_H$) values of the full sample are highlighted, showing their segregation into distinct yet overlapping regions of the parameter space.  The criterion for a strongly spatially disturbed galaxy is $A_S > 0.2$, revealing that a minority ($33\%$) of the AGN at low redshift ($z<0.1$) meet this criterion, whereas nearly all ($91\%$) AGN hosts at $z \gtrsim 0.6$ qualify.} 
\label{fig: ASvsz}
\end{figure*}

Figure~\ref{fig: ASvsz} gives an overview of the full set of galaxies in this study. The criterion for a strongly disturbed galaxy is $A_S > 0.2$, revealing that a minority ($33\%$) of the AGN at low redshift ($z<0.1$) show shape asymmetries above this value, whereas nearly all ($91\%$) AGN hosts at $z \gtrsim 0.6$ are disturbed by this criterion. This trend is similar to that for field galaxies \citep[e.g.,][]{Conselice2009}, but shows that AGN at redshifts above $z = 0.6$ have increased opportunity to grow through merger-induced disturbances. 

\subsection{Human Vision Morphology Characterization}

\begin{figure*}[h!]
   \centering
   \includegraphics[width=0.7\linewidth]{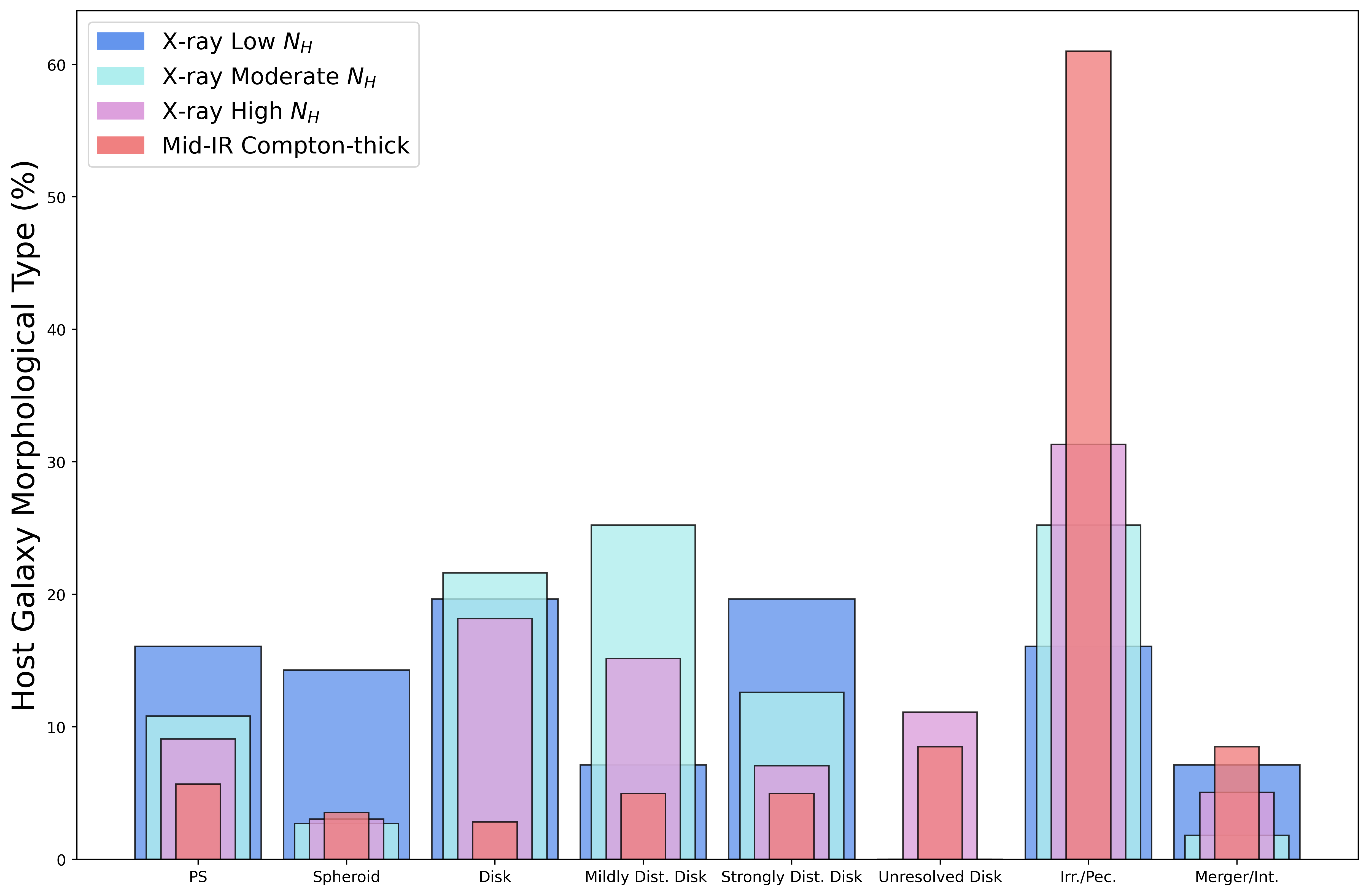}
   \centering
    \includegraphics[width=0.7\textwidth]{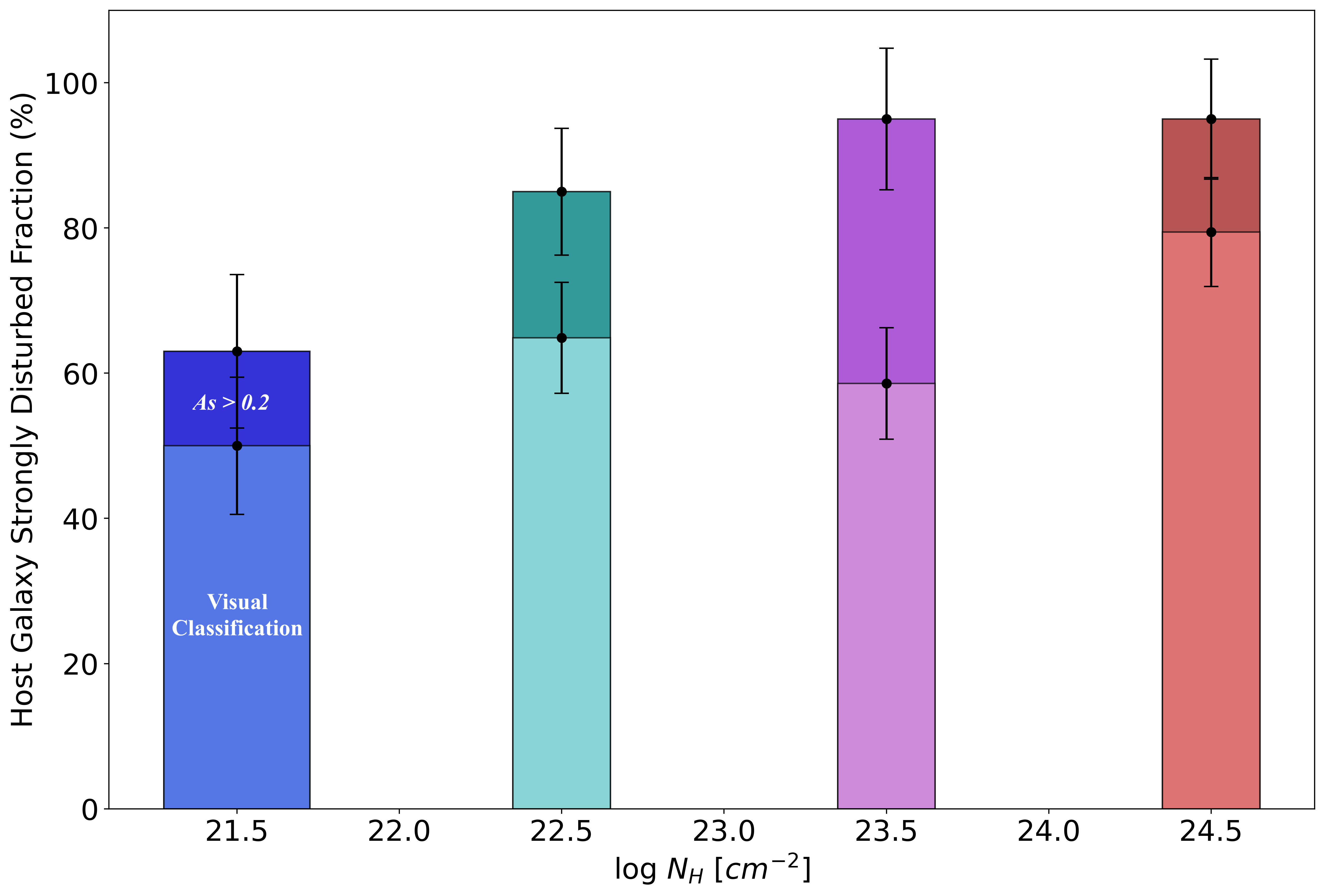}
    \caption{\textbf{Top}: The results of the visual classification of AGN host morphologies, showing the least obscured AGN dominating the undisturbed morphological categories (Point Source, Spheroid, and Disk) and the most obscured AGN, the most disturbed and disordered classes (Irregular/Peculiar) (vertical error bars denoting the $1\sigma$ binomial standard deviation are hidden for visual clarity). \textbf{Bottom}: A comparison of the results of the \emph{statmorph} $A_s\ge0.2$ (dark colors) and visual (light colors) classification methods in each $N_H$ bin, showing the total percentage of AGN host galaxies presenting as strongly asymmetric/disturbed by the two independent analyses. The bar center and width correspond to the mean and standard deviation of the $N_H$ values in each bin, respectively (simulated for the Compton-thick mid-IR-bright/X-ray-faint MIRI AGN), and the vertical error bars reflect the $1\sigma$ binomial standard deviation. It can be seen that the computer visual analysis identifies all of the strongly disturbed cases identified by the visual analysis, plus additional ones that evaded detection by the human eye. (Given that nearly all of the \emph{statmorph} $0.1<A_s<0.2$ `mild asymmetry' identifications were classified as undisturbed to the human eye, we show only the strongly disturbed \emph{statmorph} results for the most conservative comparison to the visual classification results.) }
\label{fig: vis_stat_bar_plots}
\end{figure*}

The results of the visual classification procedure for the primary AGN sample are displayed in Figure~\ref{fig: vis_stat_bar_plots}, showing the percentage of AGN host galaxies with undisturbed (`point source', `spheroid' , and `disk' ) and disturbed (`merger', `irregular', and `disturbed disk') morphologies. It can be seen that the morphology predominantly characterizing the disturbed X-ray AGN hosts is the `disturbed disk' morphology, followed by irregular; the undisturbed fraction of the X-ray AGN hosts was likewise dominated by a (symmetric) disk morphology. The MIRI-detected sample, on the other hand, appears overwhelmingly irregular/disordered, with a higher fraction of obvious mergers.   

\begin{figure*}[h!]
    \includegraphics[width=0.8\linewidth]{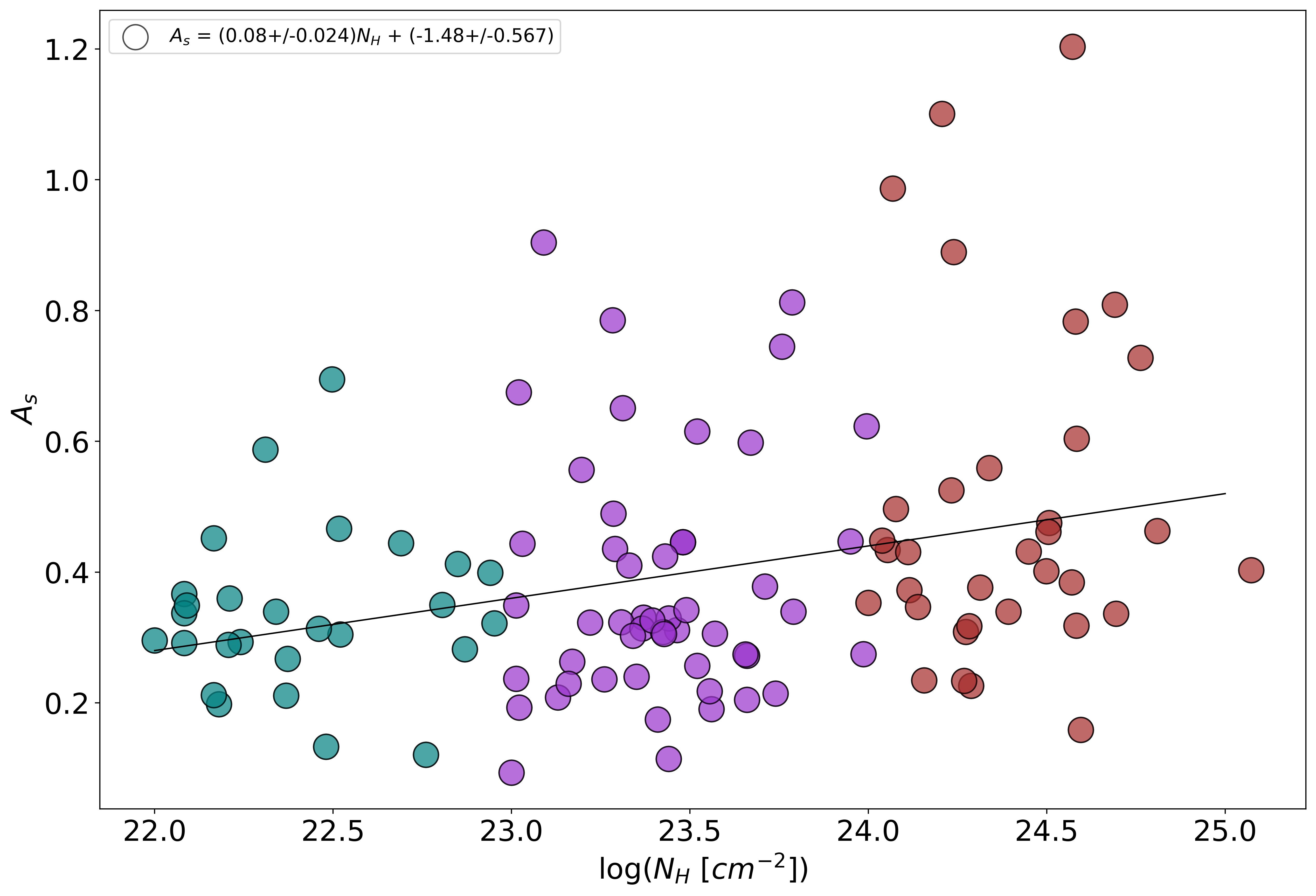}
    \centering
    \includegraphics[width=0.8\linewidth]{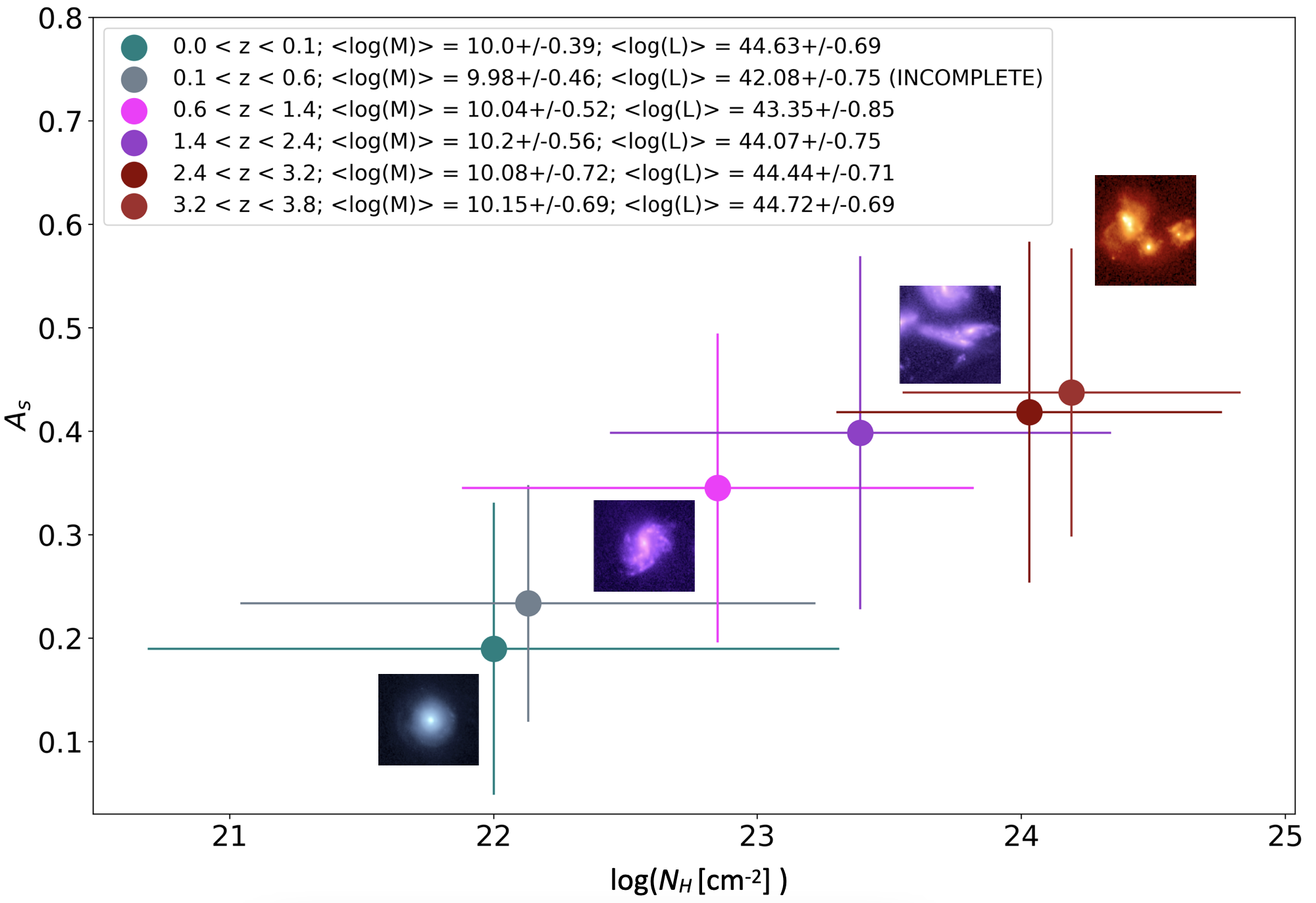}
    \caption{\textbf{Top}: Fitted \emph{statmorph} shape asymmetry ($A_{s}$) parameter values as a function of ($N_H$) obscuration for the primary sample matched on $M_{*}$, $L_{bol}$, and \textit{z} in the $N_H$ bins highlighted in colored symbols, showing a $3\sigma$ correlation. \textbf{Bottom}: Average $A_{s}$ and $N_H$ values in $L_{bol}$-\textit{z} matched bins of the full sample, shown alongside image cutouts with examples of AGN host galaxies with the corresponding shape asymmetry and obscuration level. This figure demonstrates that the dependence of $A_S$ on $N_H$ is intertwined with the dependence on redshift. At $z<0.1$, it can be seen that both lightly and highly obscured X-ray-selected AGN can show relatively low levels of disturbance; while at high redshift, there is a measurable correlation between $A_S$ and $N_H$. Overall, the linear fit to the complete data bins of the combined comparison low-redshift, and full high-redshift sample (only partially matched on source properties), reveals a $5.5\sigma$ obscuration-disturbance correlation.}
\label{fig: finalfit}
\end{figure*}

\subsection{Computer Vision Morphology Characterization}
\label{sec: statmorph_results}

In the bottom panel of Figure~\ref{fig: vis_stat_bar_plots}, we show the fraction of strongly disturbed AGN host galaxies recovered by the computational \textit{statmorph} analysis (darkly shaded bars), in comparison to the visual classification (lighter shaded bars). It is clear that this analysis method, like the visual classification procedure, also reveals a correlation between disturbed host galaxy morphology and the level of $N_H$ obscuration at $z>\sim0.6$.  The only systematic difference found between the two independent analysis methods (which proved ultimately inconsequential due to the agreement of their results) was that, in the visual classification, often what was qualitatively labeled as undisturbed, or a `mild' galaxy asymmetry/disturbance to the human eye (a subcategory `learned' after observing instances of significantly stronger disturbances), was classified by the \textit{statmorph} $A_S$ parameter as a `mild' ($A_S$=$0.1-0.2$) and `strong' ($A_S$=$0.2-0.3$) disturbance, respectively. An example of this can be seen in the two images of the same host galaxy in column 2, rows 1 and 2 of the grid of cutout images in Figure ~\ref{fig: images}. Here, the galaxy emission shown in blue in row 2 appears perfectly smooth, symmetric, and undisturbed to the unaided human eye; but in row 1, in a different SAOImage color (`aips0') that assigns a different colored contour to each level of source emission, it can be understood why \textit{statmorph} assigned this galaxy an $A_S$ value in the mildly asymmtric range: the outermost and faintest region of the galaxy emission (shown in purple) shows signs of mild disturbance. As a result, the computer vision analysis yields a higher fraction of disturbed AGN host galaxies than the visual classification procedure.  This can be attributed to the numerical precision and objectivity of the shape asymmetry algorithm over human vision in classifying objects in an inherently unbiased way, and to its enhanced sensitivity to morphological asymmetries such as faint tidal features that might go unnoticed by the human eye. However, even if we were to take a conservative approach and entirely remove the \textit{statmorph} computer vision analysis to use only visual classification, in keeping with all previous studies on the topic, our main result would still hold. In this case, we would effectively re-classify a significant number of the \emph{statmorph}-detected-but-invisible ``mildly disturbed" cases `back' to the visually undisturbed category, still leaving the ``strongly disturbed" cases that were observed using both the visual and computational methods, and that represent the bulk of the disturbed fraction.

Given the superior ability of \textit{statmorph} to detect and classify the asymmetries/disturbances of the highly obscured systems lying at the highest redshifts -- only a fraction of which were sufficiently resolved to visually distinguish between a `mild' and `strong' asymmetry (which may have led to a bias in the visual classification results where strong disturbances were more easily detected at low redshifts) -- we adopt the \textit{statmorph} analysis results as `final', and all subsequent discussion in this paper is based on them. 

\begin{figure*}[h!]

    \includegraphics[width=1\linewidth]{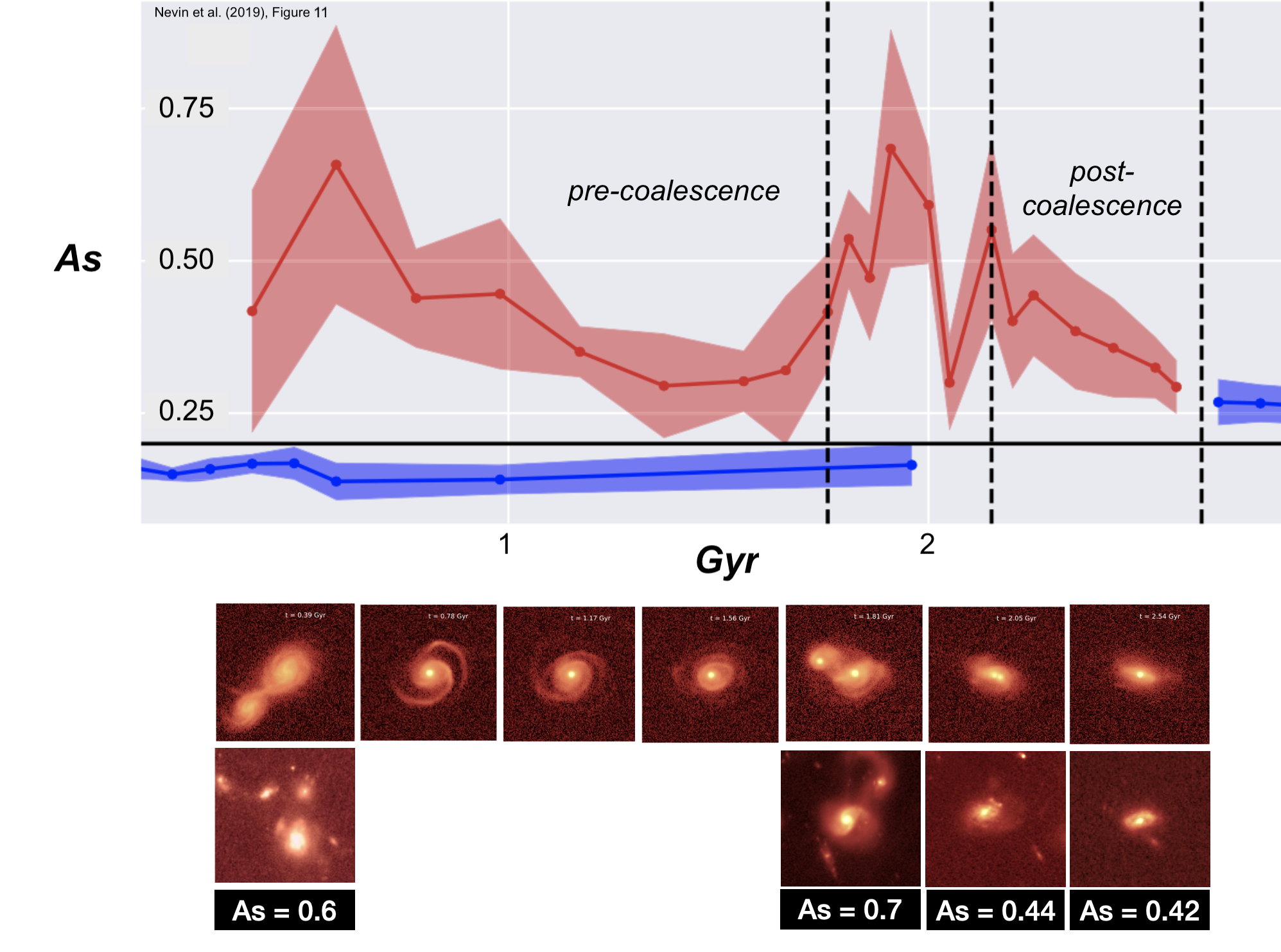}
    \centering
     \caption{\textbf{Top}: Zoom-in of lower-left panel of Figure 11 from \citet{Nevin2019} showing the evolution of the shape asymmetry parameter ($A_S$) along the major merger timeline from one of the simulations utilized in that study (non-merging galaxies are represented by the blue line). \textbf{Bottom}: Random selection of real $z\sim1.5$ JADES JWST/NIRCam F150W images of AGN host galaxies from our study (bottom row of image cutouts) classified with the same $A_S$ values as the strikingly similar simulated major-merging galaxies from Figure 2 of \citet{Nevin2019} (top row), which are shown aligned with their respective pre- and post-coalescent merger phases. This is a representative example of the power of the $A_S$ metric for identifying a range of merger signatures in galaxy imaging data over a significant portion of the merger timeline.}
\label{fig: nevin}
\end{figure*}

Considering the findings of \citet{Nevin2019}, who study the effect of each individual non-parametric morphology indicator on major and minor merger morphology identification in imaging data, we estimate the fraction of our observed AGN host disturbances that can likely be attributed to major mergers. These authors show that major and minor mergers exhibit similar $A_S$ values during different phases of their respective merger timelines, but that consideration of the \textit{G}, \textit{C}, and \textit{$M_{20}$} morphology measures together can roughly break the degeneracy. Based on their statistical findings, we estimate for the GOODS-S portion of our primary AGN sample that roughly $62\%$, $75\%$, $79\%$, and $88\%$ of the low, moderate, high, and CT subsets were induced by major mergers (see \citealp{Nevin2019} Figure 7). In Figure~\ref{fig: nevin} we show a zoom-in/stretch of the lower-left panel of their Figure 11, representing the time evolution of $A_S$ along the major merger timeline of one their simulations, where it can be seen that the $A_S$ values are characteristic of our AGN host galaxy sample. We also highlight in this figure the striking similarity between the morphologies of the JADES AGN host galaxies and those simulated in the study with the same $A_S$ values, demonstrating the robustness of this merger morphology indicator.

\subsection{The Obscuration-Disturbance Investigation}

From the analysis of 113 AGN in the primary matched sample, we find that $100\%$ are at least mildly morphologically disturbed ($A_{s}\ge0.1$), with $92\%$ satisfying the criterion for a strong, and therefore merger-induced, disturbance ($A_{s}\ge0.2$, \citealp{Pawlik2016, Nevin2019}), and $71\%$ unmistakenly identified as a strong spatial asymmetry/disturbance by the human eye (corresponding to $A_{s}\gtrsim0.3$). We measure a $3.33\sigma$ positive $N_H$-$A_S$ correlation for property-matched AGN with moderate to CT levels of obscuration, as shown in the upper plot of Figure~\ref{fig: finalfit} (reminding the reader that our sample shows an apparent true lack of low-obscuration sources above the luminosity detection threshold in the $0.6<z<2.4$ redshift range of the matched sample). Although mild, this trend indicates that the most heavily obscured AGN at z $\sim$ 2 lie in galaxies that have undergone relatively recent mergers. The lower plot in Figure~\ref{fig: finalfit} illustrates this trend in another way, where we measure the average $N_H$-$A_S$ correlation for the full sample split into redshift bins with uniform distributions of AGN luminosity, and also well-matched in average host galaxy stellar mass. Again, a trend ($5.5\sigma$) emerges in which higher values of $N_H$ are associated with larger host galaxy asymmetries. In Paper II, we attempt to disentangle the effects of the numerous interrelated parameters on this fit to the full, partially matched sample.

The analysis of the full AGN sample also reveals an apparently distinctive character in the Seyfert population at $z\ge0.6$ as compared to $z<0.1$, despite the similar selection of the two sample subsets at the same rest-frame hard X-ray energies, and their similar AGN luminosities and host galaxy stellar masses (reminding the reader of our sample incompleteness at $0.1<z<0.6$, which we aim to fill in future work). Unlike the high-redshift obscured sample, the low-redshift obscured AGN do not show a statistical preference for disturbed/merging hosts over undisturbed hosts (this does not apply however to local (U)LIRGs with hard X-ray AGN emission, see \citealp{Ricci2017}), and are nearly equally split between the two morphological classifications: of the 29/83 that are undisturbed/symmetric, 16/29 are unobscured; of the 26/83 that are mildly disturbed, 14 are unobscured; and of the 28/83 that are strongly disturbed, 12 are unobscured.  This characteristic has also been observed by others (e.g., \citealp{Koss2011, Ricci2017}). Therefore, in Paper I we only detect and quantify a statistically significant obscuration-merger connection at $z\ge0.6$. 

\section{Discussion}
\label{sec: discussion}
\subsection{Comparison to previous observational studies}

We confirm and expand upon the main conclusion of \citet{Kocevski2015}, that a majority of the heavily obscured X-ray AGN at sub-quasar luminosities are found in disks with a significant galaxy-scale spatial asymmetry/disturbance. Their work, based purely on X-ray-detected AGN, found that heavily obscured AGN are more likely to exhibit merger or interaction signatures than their non-active control counterparts at 2.5$\sigma$ significance (considering only the extended hosts). We also recover one of the main results of \citet{Donley2018}, who compared heavily obscured, IR-detected/X-ray-undetected AGN to AGN with only X-ray detection, to find that the former category are significantly more likely than the latter ($75^{+8}_{-13}$\% versus $31^{+6}_{-6}$\%) to be classified as either obviously interacting/merging, or with host morphologies so disturbed as to suggest a recent merger. The results of \citet{Donley2018} together with ours and those of \citet{Kocevski2015} provide strong evidence for high obscuration of the central AGN to be accompanied by a tendency for the host galaxy to be disturbed in a way that suggests a recent merger or similar interaction. 

\citet{Ji2022} also compared the properties of X-ray-selected AGN with IR-selected ones; the latter definition is not the same as for our study, however: they used the `power-law' selection developed by \citet{Donley2012}, which is not sensitive to the very heavily obscured, likely CT objects we classify as IR AGN. Nonetheless, their results are relevant to complement our findings: they find that the specific star formation rate (sSFR) possibly tends to be higher in the IR-selected AGN than in the X-ray ones, which we also observe conclusively amongst the AGN in our sample with $M_*\lesssim10^{10}$ $M_{\odot}$ (the results of this analysis will be presented and discussed in our forthcoming Paper II). 
Since mergers can boost SFRs, this tendency is consistent with the tendency for the most obscured AGN to lie in galaxies with increased indications for recent mergers, found in this work and in \citet{Kocevski2015} and \citet{Donley2018}. \citet{Ji2022} also find evidence that the AGN bolometric luminosity as a function of the stellar mass is higher for the IR-selected AGN than for the X-ray ones, indicating that their black holes have higher accretion rates. All of these results, although tentative, point to an evolutionary sequence for AGN from young and obscured to older and less obscured.

We additionally observe that the overall fraction of X-ray-detected AGN hosts exhibiting a disturbed morphology in rest-optical imagery grows with redshift, along with $N_H$. These two parameters are intertwined and therefore the behaviour of both $A_S$ and $N_H$ as a function of redshift will need to be carefully untangled with additional data sets, which is beyond the scope of the current analysis. In Paper II, we will consider these observed trends in a cosmological context to ascertain whether or not they are mirroring the evolving galaxy merger rate with redshift, where we expect a marked downturn in the number of galaxy mergers at $z<1$ (see, e.g., \citealp{Treister2012}, who concluded that the dominant source of SMBH growth appears to transition from secular processes locally to mergers beyond $z\sim1$, based on an analysis of ten archival surveys of moderate-luminosity AGN at $z<3$.)

\subsection{Implications of these studies}

The consensus of the studies discussed above (including the present study) is that the following trends are clearly observed within the AGN population at Cosmic Noon: (1) a very high level of significant disturbance in the host galaxies; (2) an increase in level of disturbance with increasing obscuration; 
and (3) possibly a tendency for higher AGN bolometric luminosity relative to stellar mass.  Although further work is needed to consolidate and expand on these results, in this section we speculate on their significance for AGN formation and evolution. 

Our study covers a broader range of AGN behavior than had been possible previously. By complementing our X-ray AGN sample with a newly identified sample of JWST/MIRI-detected, mid-IR-bright/X-ray-faint AGN, we probed the presumed Compton-thick regime of obscuration. Consequently, we have begun to test an evolutionary scenario for Seyferts connecting their X-ray-bright and mid-IR-bright phases. There are two aspects to this link to AGN evolution: (1) the vast majority of our primary AGN sample show strongly disturbed spatial morphologies indicative of recent merging activity; and (2) the X-ray-bright AGN in our sample present in larger numbers than the mid-IR-bright/X-ray-faint AGN, with a different dominating morphological type and broader range of AGN luminosities and obscuration levels. We expand on these points below.  

1) It is widely believed that the most luminous AGN $-$ quasars $-$ largely originate in mergers of massive galaxies \citep[e.g.,][]{Sanders1988, Sanders1996}. This hypothesis is supported by theoretical simulations \citep[e.g.,][]{Hopkins2006a, Hopkins2008, Snyder2013, Ricci2017, Kawaguchi2020} and is generally accepted in broad outline. In contrast, lower-luminosity AGN $-$ Seyferts $-$ are thought to be triggered by stochastic events in relatively passive galaxies \citep[e.g.,][]{KK2004, Treister2012, Hopkins2014}. The question is: Where is the dividing line between Seyferts and quasars, and how strict is the division? 

Host galaxy morphologies are a leading way to answer this question. An important result of our study, illustrated in Figure~\ref{fig: ASvsz}, is that for z $>$ 0.6, the great majority of AGN hosts are at least moderately disturbed ($A_S > 0.1$), and most are strongly disturbed ($A_S > 0.2$). In contrast, \citet{Hopkins2014}  conclude that a disk morphology in an AGN host should be taken as evidence of stochastic fueling by non-merger mechanisms, given the expectation that a merger would disrupt a disk and that merger-fueled AGN should only appear as bulge-dominated in a post-merger phase. However, numerous simulations  \citep[e.g.,][]{Robertson2006,Lotz2008,Hopkins2009,Darg2010,Snyder2015,Bignone2017,Nevin2019,SR2022} show that a significant fraction of spiral galaxies involved in major mergers either maintain, or relatively quickly reform, their disk morphologies after the merger event. The simulations of \citet{SR2022} show that a disk destroyed by a major merger can reform in as little as $\sim300$ Myr after the merger (see their Figure 8); and \citet{Jackson2020} show that even merging spheroids can form a gas-rich disk post-coalescence. Our findings and those of \citet{Kocevski2015} show that multiple representative samples of X-ray-selected Seyferts (i.e., modest luminosity AGN) out to Cosmic Noon show disturbed disk morphologies indicative of recent merging activity. \citet{Pierce2023} find that previous studies may have missed these features because of inadequate surface brightness depths of the observations, combined with the effects of cosmological surface brightness dimming.

Our study therefore indicates that even AGN with sub-quasar masses and luminosities are likely to experience at least one merger in their lifetime \citep[e.g.,][]{SR2022}, especially taken in context with their other physical properties that together suggest a link to the evolving cosmological merger rate\footnote{One of the observational tests of the quasar/Seyfert divide put forth in \citet{Hopkins2009b} suggests that Seyferts are not likely to be merger-induced like quasars, given that their number densities do not align with merger rates, rather that it remains relatively constant with redshift. However, we argue that, while a major gas-rich merger is necessary to produce the quasar luminosity -- and therefore we expect the quasar number density as a function of redshift to follow the evolving merger rate -- the Seyfert luminosity can be produced by alternative mechanisms. In other words, if Seyferts can be triggered by both mergers and secular processes, with the former mechanism potentially dominating at higher redshifts and the latter mechanism taking over at low redshift, we would not necessarily expect their number densities to mirror the galaxy merger rate at all redshifts.}. We observe this behavior in the realm of AGN luminosities and SED-inferred black hole masses far lower than predicted by  \citet{Hopkins2014}.  That is, mergers may play in important role in fueling AGN in the Seyfert regime.


2) Our second finding is that the observation of a trend in morphological type characterizing the X-ray-bright AGN (many disk galaxies) and mid-IR-bright/X-ray-faint AGN (mostly irregular/peculiar) suggests these two subsamples represent different AGN phases tied to a merger timeline -- especially given the likelihood that they belong to the same Seyfert AGN class as indicated by their similar host galaxy properties (e.g., radio AGN present with a distinctly different set of characteristics, as shown in \citealp{Hickox2009}). Specifically, the more disturbed morphologies 
of the hosts of the most obscured AGN suggest that they are in the early stages of evolution, just emerging from the coalescence phase.

In comparison, the larger number of X-ray-bright, less-disturbed galaxies, with a wider range in $L_{bol}$, $N_H$, and $A_S$, and confirmed to be isolated (i.e., not in a pre-coalescent merger phase), would be expected if they are a later and longer-lived phase. In other words, the X-ray-bright AGN are observed when there has been sufficient time for clearance of the CT levels of gas and dust to allow the X-ray luminosity to shine through. This evolution has only recently become evident because the most luminous, obscured, and disturbed mid-IR-bright AGN can only be detected via significant warm and hot dust re-emission in the mid-IR, as made possible by JWST.

\section{Conclusion}
\subsection{A merging population of Seyferts at Cosmic Noon}

In summary, the present study of AGN host galaxy morphologies in JWST/NIRCam imagery has supported and extended pre-JWST era observational studies that detected signs of merging activity amongst the Seyfert AGN population; as well finds a previously hidden Seyfert evolutionary sequence tied to a statistically likely major merger timeline. In addition to the availability of superior, new JWST/NIRCam imagery for more accurate measurements of rest-optical galaxy morphologies out to higher redshifts than could be accomplished previously, we attribute the robustness of our results to the following additional strengths of our analysis: 

(1) The controlled experimental design of our study, similar to the one used in \citet{Kocevski2015}, allowed for a `clean' sample that eliminated a maximum number of uncontrolled variables and observational biases that might otherwise have skewed the results of our analysis. This was achieved by partitioning a single and relatively complete sample of X-ray AGN into distinct bins of $N_H$ obscuration, all of which were detected using the same method and extracted from the same data set, and matched on numerous physical traits. We likewise cleanly extended our analysis into the CT regime of obscuration by matching a sample of mid-IR-bright/X-ray-faint AGN on the same set of galaxy properties.  In other words, the experimental design of the analysis allowed for a rigorous test of a narrowly focused scientific question: is the obscured AGN phase caused by mergers?

(2) The addition of a significant number of newly uncovered obscured AGN made possible with JWST/MIRI allowed us to extend and refine our AGN analysis in a way not possible in the pre-JWST era. Given that both mid-IR and hard X-ray wavelengths are the least sensitive to AGN obscuration, the addition of these sources to our local comparison sample of \textit{Swift}/BAT-detected AGN and \textit{Chandra} sources (selected at similar rest-frame hard X-ray energies), provided us with a complete and representative sample of Seyfert galaxies spanning multiple epochs of cosmic evolution. 

(3) The combination of human and computer visual classification methods to characterize galaxy morphologies was key in allowing us to circumvent the obstacles faced by previous authors: we observed firsthand how a non-negligible number of mildly asymmetric/disturbed disk galaxies in our sample could go unnoticed by the human eye and initially be classified as `undisturbed' -- only to discover in the results of the computer vision analysis that they are, in fact, significantly asymmetric. This was achieved through the adoption of the \emph{statmorph} shape asymmetry ($A_S$) algorithm as our primary metric for host galaxy disturbance, which is inherently unbiased in its diagnosis of galaxy morphology given that it assigns equal weights to all pixels in an input image of galaxy emission by using the associated segmentation map. While it is important to additionally consider the brightness variations amongst the pixels containing the galaxy emission in the context of different morphology studies, using the suite of classic non-parametric morphology indicators, in Paper I this would have posed as a hindrance: in this case, galaxy morphology asymmetries not linked to merging activity would have been classified as `disturbed'; and spatial disturbances truly induced by merging activity could have gone significantly undetected (the superior diagnostic power of $A_S$ over the classic asymmetry parameter, \textit{A}, in identifying mergers in imaging data is demonstrated in \citet{Nevin2019}. Furthermore, our utilization of $A_S$ as a reliable indicator of merging galaxy morphology crucially allowed us to probe a much wider window of the merger timeline than could be achieved through use of historical non-parametric measures (see \citealp{Nevin2019}), which are only effective in the relatively brief pre-coalescent and active merger phases (see., e.g., \citealp{Lotz2008} and \citealp{Bignone2017}). This suggests that a majority of previous merger studies may not have characterized the post-merger remnant phase.

\subsection{Foundation for future work}

Our confirmation of a correlation between the level of AGN host galaxy disturbance and AGN obscuration amongst a complete, multiwavelength sample of obscured Seyfert AGN at $z=0-4$, implies that the obscured Seyfert AGN phase may mark a period of significant supermassive black hole (SMBH) growth, mirroring the quasar population defined by higher black hole masses and AGN luminosities. This hypothesis is supported by \citet{McAlpine2020}, a study that utilizes the cosmological hydrodynamical EAGLE simulation to investigate the degree to which black hole (BH) activity is enhanced during the period of a major merger. They find that the galaxies embodying the most optimal conditions for triggering an AGN via a merger have the relatively low stellar masses, black hole masses, and luminosities typical of the AGN in our sample. The second key discovery of their study is that 50-75\% of major-merger-triggered BH activity occurs in post-merger systems, \textit{after} the two galaxies have coalesced. This implies that the majority of observational merger studies, which focus on the earliest stages of a merger, may have missed a significant fraction of the peak of BH growth in their samples, and mistakenly classified the post-merger remnants as isolated, non-merger AGN. 

Furthermore, the recovery of a physical distinction between the obscured X-ray-bright AGN and the mid-IR-bright/X-ray-faint extension into the CT regime, suggests, in the context of their other physical properties, that these two subsets may mark distinct phases of an evolutionary scenario tied to a major merger timeline. We investigate this observed trend further in Paper II; and also consider that the definition of a major merger appears somewhat arbitrary in theoretical simulations. For example, the cosmological model of \citet{Volonteri2009} uses a 1:10 mass ratio threshold to define a major major, which is consistent with widespread galaxy merging leading to the establishment of the M-$\sigma$ relation. This definition, however, is updated to 3:10 in \citet{Volonteri2013} to accommodate alternative fueling mechanisms for AGN. This definition is important to maintain consistency between theory and observations as they are improved, and the role of major mergers in AGN formation is better defined. 
   
The strength of our conclusion lies not only in the corroboration of previous AGN host morphology studies and expansion to higher redshifts and levels of AGN obscuration, but also in the agreement with the various merger simulations discussed. In Paper II, we will expand upon the analysis of this paper with supplementary data from the literature to further investigate if the Seyfert population sampled may evolve according to a lower-luminosity, lower-mass extension of the accepted quasar evolution model, which is based almost exclusively on major mergers.

\clearpage

\appendix

Here we show JADES v1 NIRCam mosaic cutout images of a randomly chosen selection of AGN in our sample at the effective wavelengths available (listed in the image caption), to demonstrate the progressive resolution degradation inevitably caused by an increasing PSF size towards longer wavelengths. This instrumental effect motivated our decision to include only the F150W mosaic imagery in our analysis, as it represents the mosaic at/near peak NIRCam resolution and sensitivity that also maximizes the coverage of our AGN sample in the GOODS-S field. We observed, after repeatedly visually examining a statistically significant number images of the AGN host galaxies in a number of trials, that no information would be gained or lost, or conclusions altered, by measuring the AGN host galaxy morphology at a constant rest-frame wavelength, requiring our analysis procedure to be repeated using mosaic image cutouts at a range of observed wavelengths other than (longward of) F150W. It can be witnessed here in all images shown, that the morphological identification of each galaxy as measured in the F150W image as symmetric/undisturbed or asymmetric/disturbed, remains unchanged in the images exposed at other wavelengths; all galaxy images exposed at wavelengths longward of F150W appear identical to the galaxy image at F150W, except at poorer resolution. We also note a similar observance of the images taken at wavelengths shortward of F150W: while they appear equally or slightly more resolved than the F150W image (but noisier), the galaxy morphology appears identical and therefore no new information would be gleaned from inclusion of these images in our analysis.

  \begin{figure}
     \centering
         \centering
         \includegraphics[width=1\linewidth]{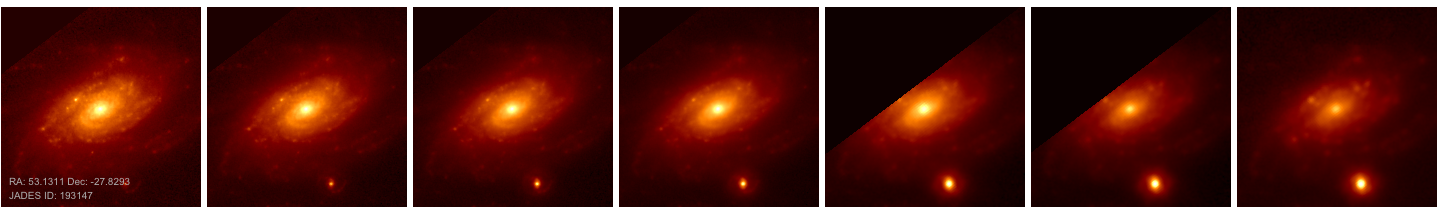}
         \centering
         \includegraphics[width=1\linewidth]{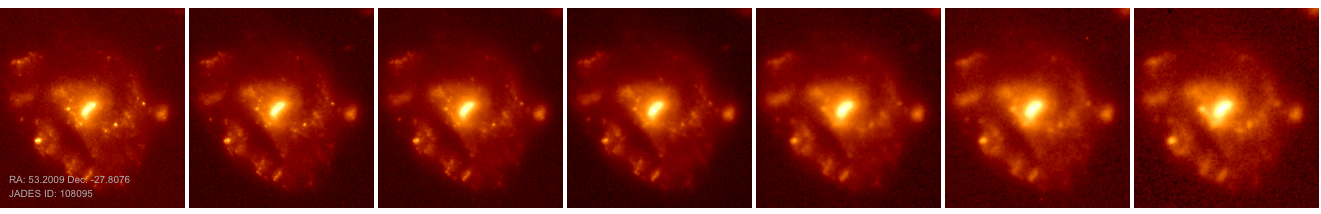}   
         \centering
         \includegraphics[width=1\linewidth]{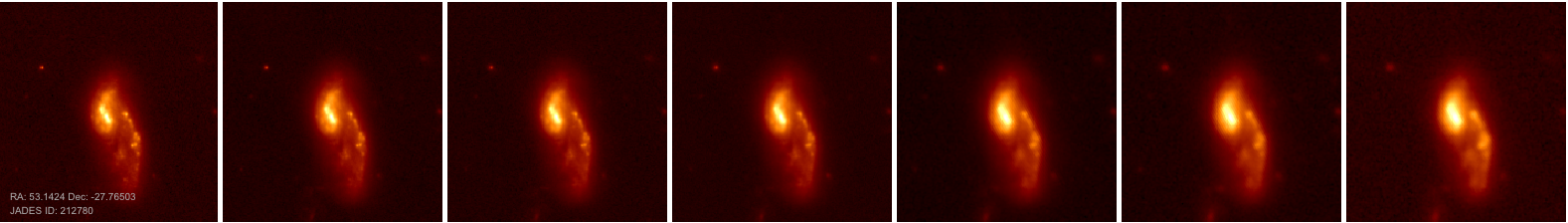}
         \centering
         \includegraphics[width=1\linewidth]{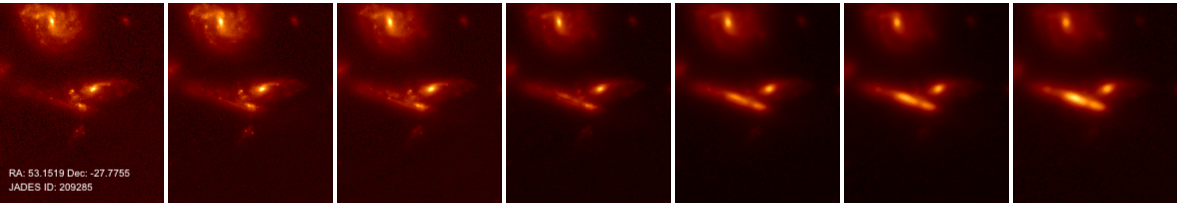}
         \centering
         \includegraphics[width=1\linewidth]{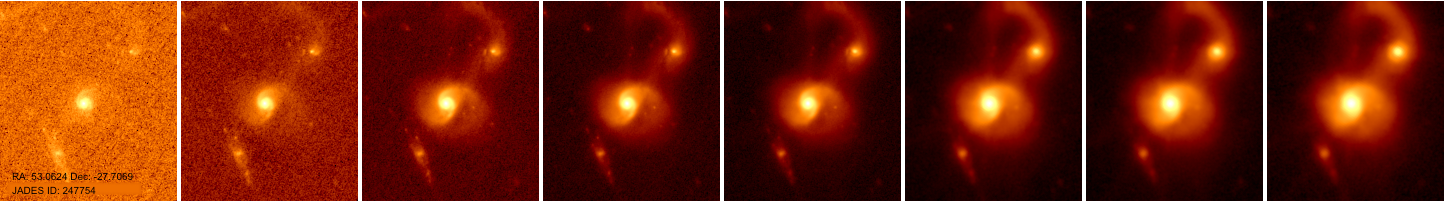}
         \centering
         \includegraphics[width=1\linewidth]{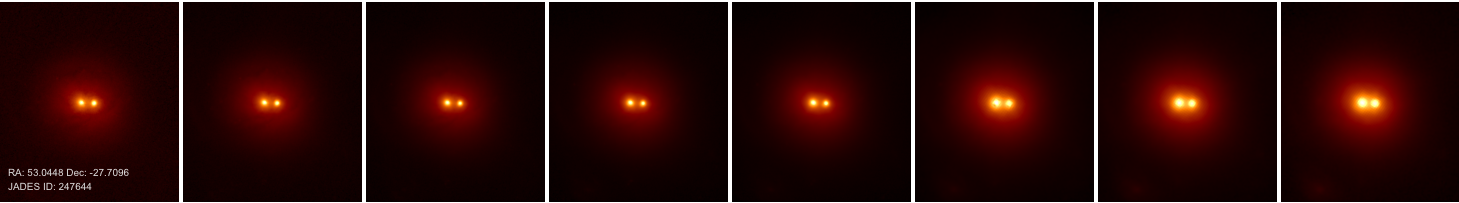}
         
         \caption{Example JADES v1 NIRCam GOODS-S mosaic image cutouts shown in a variety of the wide filters, highlighting the progressive resolution degradation inevitably caused by a growing PSF size towards longer wavelengths. Here it can be seen that the F150W band chosen for our AGN host galaxy morphology analysis offers maximal resolution and signal-to-noise in comparison to the other bands available (as well as maximizes the coverage of the chosen AGN sample; see Section ~\ref{sec: nircam} for further details). Rows 1-4 show images in the following NIRCam bands: F090W, F115W, F150W, F200W, F277W, F356W, F444W; and rows 5-6: F070W, F090W, F115W, F150W, F200W, F277W, F356W, F444W. The top three rows show lightly obscured sources ($N_H<10^{22} cm^{-2}$) at $z=0.2-0.3$; the fourth row shows a highly obscured source ($N_{H} = 10^{23}$ $cm^{-2}$) at $z=1$; the fifth row shows a lightly obscured source ($N_{H} = 10^{21.5}$ $cm^{-2}$) at $z=2$; and the sixth row shows a moderately obscured source ($N_{H} = 10^{22.1}$ $cm^{-2}$) at $z=0.4$.}
    
  \label{fig: appendix}
  \end{figure}

\vspace{5mm}

\clearpage

\begin{acknowledgments}
This work was supported by NASA grants NNX13AD82G,  1255094, and the NASA contract for NIRCam to the University of Arizona, NAS5-02015. This work is based on observations made with the NASA/ESA/CSA James Webb Space Telescope. SA, NB, DJE, KH, ZJ, BR, MR, and CNAW acknowledge support from the NIRCam Science Team contract to the University of Arizona, NAS5-02015.  DJE is further supported as a Simons Investigator. SA, JL, IS, and GHR acknowledge support from the JWST Mid-Infrared Instrument (MIRI) Science Team Lead, grant 80NSSC18K0555, from NASA Goddard Space Flight Center to the University of Arizona. These observations are associated with JWST GTO programs \#1180 and \#1207, and GO program \#1963. AJB acknowledges funding from the ``FirstGalaxies" Advanced Grant from the European Research Council (ERC) under the European Union's Horizon 2020 research and innovation program (Grant agreement No. 789056). The work of CCW is supported by NOIRLab, which is managed by the Association of Universities for Research in Astronomy (AURA) under a cooperative agreement with the National Science Foundation. WB and FDE acknowledge support by the Science and Technology Facilities Council (STFC) and by the ERC through Advanced Grant 695671 "QUENCH".

The JWST/NIRCam version 1 mosaic images presented and analyzed in this article were obtained internally from within the JADES team, before the data became public. The (unaltered) data products have since been made public and available for download from both the JADES Public Data Access webpage (https://jades-survey.github.io/scientists/data.html), as well as the Mikulski Archive for Space Telescopes (MAST) at the Space Telescope Science Institute. The specific observations analyzed can be accessed via \dataset[doi: 10.17909/8tdj-8n28]{https://doi.org/10.17909/8tdj-8n28}.
\end{acknowledgments}
\facility{JWST}

\bibliographystyle{aasjournal}

\end{document}